\documentclass[12pt,letterpaper]{article}
\usepackage{epsfig,rotating,setspace,latexsym,amsmath,epsf,amssymb,amsfonts,bm,theorem,cite,caption,subcaption,enumerate,longtable,accents}
\usepackage{algorithm,algorithmic,graphicx,epsf,authblk,epstopdf,url,color,multirow}
\newtheorem{theorem}{Theorem}

\newtheorem{corollary}{Corollary}

\newtheorem{remark}{Remark}
\newtheorem{lemma}{Lemma}
\newenvironment{Proof}[1]{\medskip\par\noindent{\bf Proof:\,}\,#1}{{\mbox{\,$\blacksquare$}\par}}

\newcommand{\cs}{{\mathcal{S}}}
\newcommand{\bt}{{\boldsymbol{\tau}}}
\newcommand{\bc}{{\mathbf{C}}}
\newcommand{\st}{{\text{s.t.}}}

\allowdisplaybreaks

\begin{document}
	
\title{Noisy Private Information Retrieval: On Separability of Channel Coding and Information Retrieval\thanks{This work was supported by NSF Grants  CNS 13-14733, CCF 14-22111, CNS 15-26608 and CCF 17-13977.}}
	
\author{Karim Banawan \qquad Sennur Ulukus\\
	\normalsize Department of Electrical and Computer Engineering\\
	\normalsize University of Maryland, College Park, MD 20742 \\
	\normalsize {\it kbanawan@umd.edu} \qquad {\it ulukus@umd.edu}}
	
\maketitle
	
\vspace*{-0.5cm}

\begin{abstract}
	We consider the problem of noisy private information retrieval (NPIR) from $N$ non-communicating databases, each storing the same set of $M$ messages. In this model, the answer strings are not returned through noiseless bit pipes, but rather through \emph{noisy} memoryless channels. We aim at characterizing the PIR capacity for this model as a function of the statistical information measures of the noisy channels such as entropy and mutual information. We derive a general upper bound for the retrieval rate in the form of a max-min optimization. We use the achievable schemes for the PIR problem under asymmetric traffic constraints and random coding arguments to derive a general lower bound for the retrieval rate. The upper and lower bounds match for $M=2$ and $M=3$, for any $N$, and any noisy channel. The results imply that separation between channel coding and retrieval is optimal except for adapting the traffic ratio from the databases. We refer to this as \emph{almost separation}. Next, we consider the private information retrieval problem from multiple access channels (MAC-PIR). In MAC-PIR, the database responses reach the user through a multiple access channel (MAC) that mixes the responses together in a stochastic way. We show that for the additive MAC and the conjunction/disjunction MAC, channel coding and retrieval scheme are \emph{inseparable} unlike in NPIR. We show that the retrieval scheme depends on the properties of the MAC, in particular on the linearity aspect. For both cases, we provide schemes that achieve the full capacity without any loss due to the privacy constraint, which implies that the user can exploit the nature of the channel to improve privacy. Finally, we show that the full unconstrained capacity is not always attainable by determining the capacity of the selection channel.  
\end{abstract}

\section{Introduction}
In the era of big data, efficient data-mining techniques are present everywhere, from social media to online-shopping and search history. These new challenges motivate studying the privacy issues that arise in modern networks. Private information retrieval (PIR), introduced by Chor et al. \cite{ChorPIR} and remained an important research avenue in computer science community (see for example\cite{ChorPIR, PIRsurvey2004, cachin1999computationally, ostrovsky2007survey, yekhanin2010private}), is a canonical problem to study the privacy of the downloaded content from public databases. In the classical PIR, a user wishes to retrieve a file privately from $N$ distributed and non-colluding databases each storing the same set of $M$ messages (files), in a way that no database can learn the identity of the user's desired file. To that end, the user submits queries for the databases that do not reveal the user's interest in the desired file. The databases respond with \emph{correct} answer strings via \emph{noiseless orthogonal links}, from which the user reconstructs the desired file. PIR schemes are designed to be more efficient than the trivial scheme of downloading all the files stored in the databases in terms of the retrieval rate, which is defined as the ratio between the number of downloaded bits from the desired message and the total download.

Recently, the PIR problem has attracted a renewed interest within the information theory community \cite{RamchandranPIR, unsynchonizedPIR, YamamotoPIR, VardyConf2015, RazanPIR}. In order to characterize the fundamental limits of the problem, Sun-Jafar introduced the notion of PIR capacity $C_{\text{PIR}}$ in \cite{JafarPIR}, which is defined as the supremum of all PIR rates over all achievable retrieval schemes.  \cite{JafarPIR} proved that for the classical PIR model, $C_{\text{PIR}}=(1+\frac{1}{N}+\cdots+\frac{1}{N^{M-1}})^{-1}$. The achievability scheme is a greedy algorithm that employs a \emph{symmetric query} structure for all databases. Following \cite{JafarPIR}, the capacities of many interesting variants of the classical PIR problem have been considered \cite{JafarColluding, symmetricPIR, KarimCoded, arbmsgPIR, codedsymmetric, MultiroundPIR, codedcolluded, codedcolludedJafar, arbitraryCollusion, MPIRjournal, codedcolludingZhang, MPIRcodedcolludingZhang, BPIRjournal,RobustPIR_Razane, symmetricByzantine, tandon2017capacity, wang2017linear, kadhe2017private, wei2017fundamental, chen2017capacity, wei2017capacity, SecurePIR, sun2017_computation, mirmohseni2017private, abdul2017private, wei2017fundamental_partial,KarimAsymmetricPIR,PIR_WTC_II,PrivateSearch,CodeColludeByzantinePIR, StorageConstrainedPIR, StorageConstrainedPIR_Wei}.

In all previous works, the links from the databases to the user are assumed to be noiseless. Furthermore, these works assume that the answer strings are returned via orthogonal links, i.e., the user receives $N$ separate answer strings, which are not mixed. There are many practical settings where these assumptions may not be valid. For instance, while browsing (retrieving information on) the internet, some packets may be dropped randomly. This scenario can be abstracted out as passing the answer strings through an erasure channel. Alternatively, the data packets may be randomly corrupted, which can be modeled as a binary symmetric channel that flips randomly some symbols in the answer strings. Consequently, a more realistic realistic retrieval model may be to assume that the databases return their answer strings through memoryless noisy channels with known transition probabilities. The noisy nature of the channel induces random errors along the received answer strings. 

Yet, in other applications, the  answer strings may be mixed before reaching the user. For example: if the user is retrieving the desired file from wireless base stations, the answer strings would be combined on the air before reaching the user. Another example is retrieval from a cloud, where the returned packets may collide and superimpose each other. These practical settings can be represented with another abstract model, which is the cooperative multiple access channel (MAC) model, where the databases cooperate to convey the desired message to the user, while the user receives a stochastic mapping from the database responses in general. These two cases, namely, \emph{noisy} and \emph{multiple access} nature of retrieval channels, pose many interesting questions, such as: How to devise schemes that mitigate the errors introduced by the channel with a small sacrifice from the private retrieval rate? Is there a \emph{separation} between the channel coding needed for reliable transmission over noisy channels and the private retrieval scheme, or if there is a necessity for joint processing? How do the statistical properties of the noisy channels fundamentally affect the private retrieval rate?

In this paper, we introduce noisy PIR with orthogonal links (NPIR) and PIR problem from multiple access channel (MAC-PIR).  We first focus on the NPIR problem and then consider the MAC-PIR problem in Section~\ref{MAC_PIR_sec}. In NPIR, the $n$th database is connected to the user via a discrete memoryless channel with known transition probability distribution $p(y_n|x_n)$. Hence, the user needs to decode the desired message \emph{reliably} by observing the noisy versions of the returned answer strings. Intuitively, since a channel with worse channel condition needs a lower code rate to combat the channel errors, we do not expect the lengths of the answer strings to be the same from all the databases. Therefore, in this work, we allow the traffic from each database to be \emph{asymmetric} as in \cite{KarimAsymmetricPIR} and \cite{PIR_WTC_II}. In this work, we aim at characterizing the capacity of the NPIR problem in terms of the statistical information measures of the noisy channels such as mutual information, the number of messages $M$, and the number of databases $N$. To that end, we first derive a general upper bound for the retrieval rate in the form of a max-min problem. The converse proof is inspired by the converse proof in \cite{KarimAsymmetricPIR}, in particular in the way the asymmetry is handled. We show the achievability proof by random coding arguments and enforcing the uncoded responses to operate at one of the corner points of the PIR problem under asymmetric traffic constraints. The upper and lower bounds match for $M=2$ and $M=3$ messages, for arbitrary $N$ databases, and any noisy channel. Our results show that the channel coding needed to mitigate the channel errors and the retrieval scheme are \emph{almost separable} in the sense that the noisy channels affect only the traffic ratio requested from each database and not the explicit coding technique. Interestingly, the upper and lower bounds depend only on the \emph{capacity} of the noisy channels and not on the explicit transition probability of the channels.

In the MAC-PIR problem, the responses of the databases reach the user through a discrete memoryless MAC with a known transition probability $p(y|x_1, \cdots, x_N)$. In this case, the output of the channel is a mixture (possibly noisy mixture) of all databases responses. The user needs to decode the desired message with vanishingly small probability of error from the output of the channel. Interestingly, for this model, we show that channel coding and retrieval strategy are \emph{inseparable} unlike in the NPIR problem. We show this fact by deriving the PIR capacity of two simple MACs, namely: additive MAC, and logical conjunction/disjunction MAC. In these two cases, we show that \emph{privacy for free} can be attained by designing retrieval strategies that exploit the properties of the channel to maximize the retrieval rate. Interestingly, we show that for the additive MAC, the optimal PIR scheme is linear, while for the logical conjunction/disjunction MAC we show that a non-linear PIR scheme, that requires $N \geq 2^{M-1}$ is needed to achieve $C_{PIR}=1$. We conclude this discussion by showing that full unconstrained capacity may not be attainable for all MACs by giving a counterexample, which is the selection MAC, which has a capacity of $C_{PIR}=\frac{1}{M}$. The exact PIR capacity of the MAC-PIR for an arbitrary transition probability distribution remains an open problem in general.

\section{System Model}\label{classicalPIR}
We consider a classical PIR model with $N$ replicated and non-communicating databases storing $M$ messages. Each database stores the same set of messages $W_{1:M}=\{W_1, \cdots, W_M\}$. The $m$th message $W_m$ is an $L$-length binary (without loss of generality) vector picked uniformly from $\mathbb{F}_2^L$. The messages $W_{1:M}$ are independent and identically distributed, i.e.,
\begin{align}
H(W_m)=&L, \quad m \in \{1, \cdots, M\} \\
H(W_{1:M})=&ML \label{msg_indep}
\end{align}

In PIR, a user wants to retrieve a message $W_i$ reliably and privately. To that end, the user submits $N$ queries $Q_{1:N}^{[i]}=\{Q_1^{[i]}, \cdots, Q_N^{[i]}\}$, one for each database. Since the user does not have any information about the message set in advance, the queries and the messages are statistically independent, 
\begin{align}\label{independency}
I(W_{1:M};Q_{1:N}^{[i]})=0, \quad i \in \{1, \cdots, M\}
\end{align} 
The $n$th database responds to $Q_n^{[i]}$ with a $t_n$-length answer string $A_n^{[i]}=(X_{n,1}^{[i]}, \cdots, X_{n,t_n}^{[i]})$.  The $n$th answer string is a deterministic function of the messages $W_{1:M}$ and the query $Q_n^{[i]}$, hence,
\begin{align}\label{answer_constraint}
H(A_n^{[i]}|W_{1:M},Q_n^{[i]})=0, \quad n \in \{1, \cdots, N\}, \:\: i \in \{1, \cdots, M\}
\end{align}

\begin{figure}[t]
	\centering
	\includegraphics[width=1\textwidth]{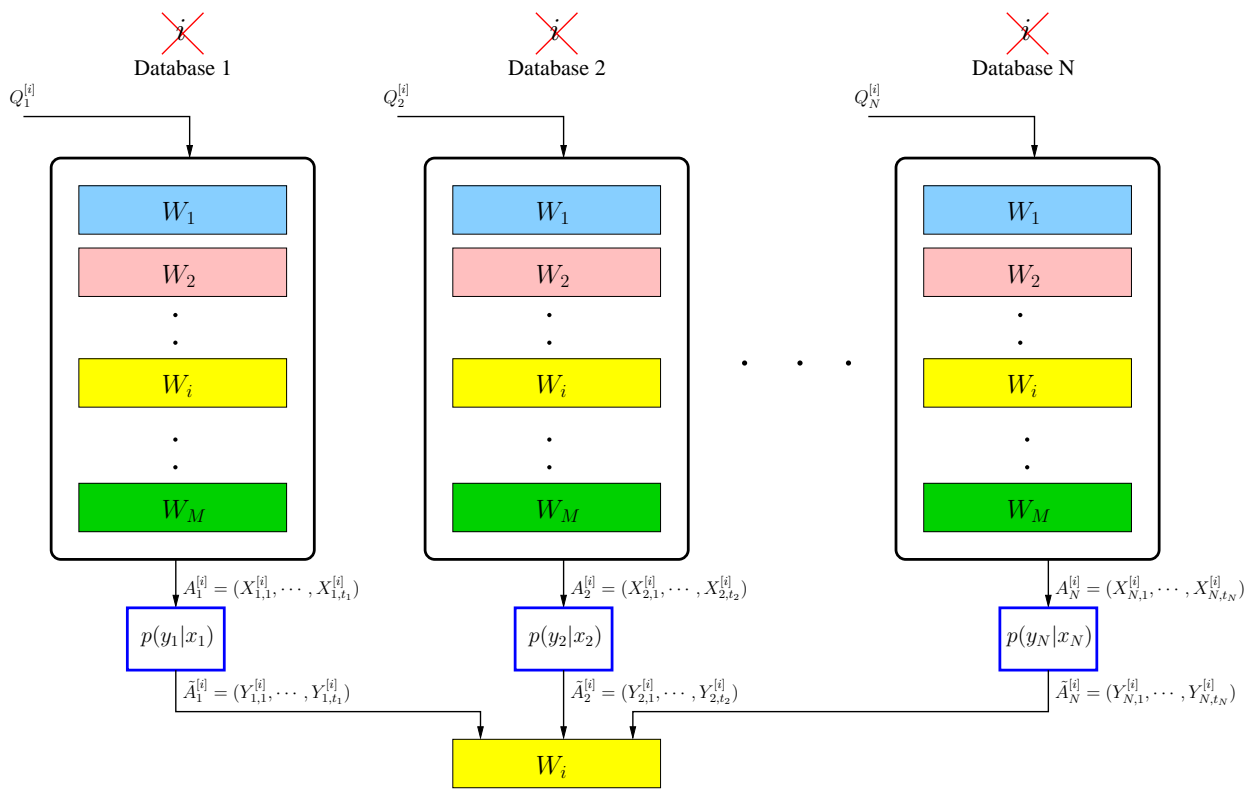}
	\caption{The noisy PIR (NPIR) problem.}
	\label{PIR_noisy}
	\vspace*{-0.4cm}
\end{figure}

In noisy PIR with orthogonal links (NPIR, see Fig.~\ref{PIR_noisy}), the user receives the $n$th answer string via a discrete memoryless channel (response channel) with a transition probability $p(y_n|x_n)$. In this model, the noisy channels are \emph{orthogonal}, in the sense that the noisy answer strings do not interact (mix). Thus, the user receives a noisy answer string $\tilde{A}_n^{[i]}=(Y_{n,1}^{[i]}, \cdots, Y_{n,t_n}^{[i]})$. Therefore, we have, 
\begin{align}\label{memoryless}
P\left(\tilde{A}_n^{[i]}=(y_{n,1}^{[i]}, \cdots, y_{n,t_n}^{[i]})|A_n^{[i]}=(x_{n,1}^{[i]}, \cdots, x_{n,t_n}^{[i]})\right)=\prod_{\eta_n=1}^{t_n} p\left(y_{n,\eta_n}^{[i]}|x_{n,\eta_n}^{[i]}\right)
\end{align}
Consequently, $(W_{1:M},Q_{n}^{[i]}) \rightarrow A_n^{[i]} \rightarrow \tilde{A}_n^{[i]}$ forms a Markov chain. Let us denote the channel capacity of the $n$th response channel by $C_n$, denote,
\begin{align}
C_n=\max_{p(x_n)} \:\: I(X_n;Y_n)
\end{align}
where $X_n$, $Y_n$ are the single-letter input and output pair for the $n$th response channel. Without loss of generality, assume that the channel capacities are ordered such that $C_1 \geq C_2 \geq \cdots \geq C_N$, i.e., the channel capacities form a non-increasing sequence. Let $\bc=(C_1, \cdots, C_N)$ be the vector of the channel capacities.  

We note that, in general, the user and the databases can agree on suitable lengths $\{t_n\}_{n=1}^N$ for the answer strings, which may not be equal in general, such that they maximize the retrieval rate. Let us define the traffic ratio vector $\bt=(\tau_1, \cdots, \tau_N)$ as,
\begin{align}\label{traffic_ratio}
\tau_n=\frac{t_n}{\sum_{j=1}^{N} t_j}, \quad n \in \{1, \cdots, N\}
\end{align}

To ensure privacy, the queries $Q_{1:N}^{[i]}$ should be designed such that the query to the $n$th database does not reveal any information about $i$. We can write the privacy constraint as   
\begin{align}\label{privacy_constraint}
(Q_n^{[i]},A_n^{[i]},W_{1:M}) \sim (Q_n^{[j]},A_n^{[j]},W_{1:M}), \quad \forall i,j \in \{1, \cdots, M\} 
\end{align}
We note that from privacy constraint and due to the Markov chain $(W_{1:M},Q_{n}^{[i]}) \rightarrow A_n^{[i]} \rightarrow \tilde{A}_n^{[i]}$, we may write that $(Q_n^{[i]},A_n^{[i]},\tilde{A}_n^{[i]},W_{1:M},W_{1:M}) \sim (Q_n^{[j]},A_n^{[j]},\tilde{A}_n^{[j]},W_{1:M}), \quad \forall i,j \in \{1, \cdots, M\}$.     
 
In addition, the user should be able to reconstruct the desired message $W_i$ by observing the noisy answer strings $\tilde{A}_{1:N}^{[i]}$ with arbitrarily small probability of error $P_e(L)$, i.e., $P_e(L) \rightarrow 0$ as $L \rightarrow \infty$. Hence, from Fano's inequality, we have,

\begin{align}\label{reliability_constraint}
H(W_i|Q_{1:N}^{[i]}, \tilde{A}_{1:N}^{[i]}) \leq 1+P_e(L) \cdot L =o(L)
\end{align}
where $\frac{o(L)}{L} \rightarrow 0$ as $L \rightarrow \infty$.

For a fixed traffic ratio vector $\bt$, the retrieval rate $R(\bt,\bc)$ is achievable if there exists a sequence of retrieval schemes, indexed by the message length $L$, that satisfy the privacy constraint \eqref{privacy_constraint} and the reliability constraint \eqref{reliability_constraint} with answer string lengths $\{t_n\}_{n=1}^N$ that conform with \eqref{traffic_ratio}, thus,
\begin{align}
R(\bt,\bc)=\lim_{L \rightarrow \infty} \frac{L}{\sum_{n=1}^{N} t_n}
\end{align}

Consequently, the retrieval rate $R(\bc)$ is the supremum of $R(\bt,\bc)$ over all traffic ratio vectors in $\mathbb{T}=\{(\tau_1, \cdots, \tau_N): \tau_n \geq 0 \:\: \forall n, \sum_{n=1}^{N} \tau_n=1\}$. The PIR capacity for this model $C_{\text{PIR}}(\bc)$ is given by
\begin{align}
C_{\text{PIR}}(\bc)=\sup\:\: R(\bc)
\end{align} 
where the supermum is over all achievable retrieval schemes.
\section{Main Results and Discussions on NPIR}
In this section, we present the main results of the NPIR problem. The first result gives an upper bound for the NPIR problem.

\begin{theorem}[Upper bound]\label{Thm1}
	For NPIR with noisy links of capacities $\bc=(C_1, \cdots,C_N)$, the retrieval rate is upper bounded by,
	\begin{align}\label{upper_bound}
	C_{\text{PIR}}(\bc) \leq \bar{C}_{\text{PIR}}(\bc)=\max_{\boldsymbol{\tau} \in \mathbb{T}} \min_{n_i \in \{1, \cdots, N\}}\!\!\!\! \frac{\sum_{n=1}^N \tau_n C_n+\frac{\sum_{n=n_1+1}^N \tau_n C_n}{n_1}+\cdots+\frac{\sum_{n=n_{M-1}+1}^N \tau_n C_n}{\prod_{i=1}^{M-1} n_i}}{1+\frac{1}{n_1}+\cdots+\frac{1}{\prod_{i=1}^{M-1}n_i}}
	\end{align}
	where $\mathbb{T}=\left\{\boldsymbol{\tau}: \tau_n \geq 0 \quad \forall n \in [1:N],\quad \sum_{n=1}^{N} \tau_n=1\right\}$.
\end{theorem}

The proof of this upper bound is given in Section~\ref{converse}. The second result gives an achievability scheme for the NPIR problem.

\begin{theorem}[Lower bound]\label{Thm2}
	For NPIR with noisy links of capacities $\bc=(C_1, \cdots, C_N)$, for a monotone non-decreasing sequence $\mathbf{n}=\{n_i\}_{i=0}^{M-1} \subset \{1, \cdots, N\}^{M}$, let $n_{-1}=0$, and $\cs=\{i \geq 0: n_i-n_{i-1}>0\}$. Denote $y_\ell[k]$ to be the number of stages of the achievable scheme that downloads $k$-sums from the $n$th database in one repetition of the scheme, such that $n_{\ell-1} \leq n \leq n_{\ell}$, and $\ell \in \cs$. Let $\xi_\ell=\prod_{s \in \cs \setminus \{\ell\}} \binom{M-2}{s-1}$. The number of stages $y_\ell[k]$ is characterized by the following system of difference equations:
	\begin{align}\label{differenceEqn}
	y_0[k]&=(n_0\!-\!1)y_0[k\!-\!1]+\sum_{j \in \cs \setminus \{0\}} (n_j\!-\!n_{j-1}) y_j[k\!-\!1] \notag\\
	y_1[k]&=(n_1\!-\!n_0\!-\!1)y_1[k\!-\!1]+\sum_{j \in \cs \setminus \{1\}} (n_j\!-\!n_{j-1}) y_j[k\!-\!1] \notag\\
	y_\ell[k]&=n_0 \xi_\ell \delta[k\!-\!\ell\!-\!1]+(n_\ell\!-\!n_{\ell-1}\!-\!1) y_\ell[k-1]+\sum_{j \in \cs \setminus \{\ell\}} (n_j\!-\!n_{j-1})y_j[k\!-\!1], \quad  \ell \geq 2
	\end{align} 
	where $\delta[\cdot]$ is the Kronecker delta function. The initial conditions of \eqref{differenceEqn} are $y_0[1]=\prod_{s \in \cs} \binom{M-2}{s-1}$, and $y_j[k]=0$ for $k \leq j$. 
	Then, the achievable rate corresponding to $\mathbf{n}$ is given by:
	\begin{align}
	R(\mathbf{n},\bc)=\frac{\sum_{\ell \in \cs} \sum_{k=1}^{M}\binom{M-1}{k-1} y_\ell[k](n_\ell-n_{\ell-1})}{\sum_{\ell \in \cs} \sum_{n=n_{\ell-1}+1}^{n_\ell}\frac{\sum_{k=1}^{M} \binom{M}{k} y_\ell[k] }{C_n}}
	\end{align}
	Consequently, the capacity $C_{\text{PIR}}(\bc)$ is lower bounded by:
	\begin{align}
	C_{\text{PIR}}(\bc) \geq R(\bc)&=\max_{n_0 \leq \cdots \leq n_{M-1} \in \{1, \cdots, N\}} R(\mathbf{n},\bc)\\
	&=\max_{n_0 \leq \cdots \leq n_{M-1} \in \{1, \cdots, N\}} \frac{\sum_{\ell \in \cs} \sum_{k=1}^{M}\binom{M-1}{k-1} y_\ell[k](n_\ell-n_{\ell-1})}{\sum_{\ell \in \cs} \sum_{n=n_{\ell-1}+1}^{n_\ell}\frac{\sum_{k=1}^{M} \binom{M}{k} y_\ell[k] }{C_n}}
	\end{align} 
\end{theorem}

The proof of this lower bound is given in Section~\ref{achievability}. We have the following remarks.

\begin{remark}
	The upper and lower bounds for the retrieval rate are similar to the corresponding bounds for the PIR-WTC-II problem \cite{PIR_WTC_II} after replacing the secrecy capacity of WTC-II, $1-\mu_n$, with the capacity of the noisy link $C_n$. Thus, the NPIR problem inherits all the structural remarks of the PIR-WTC-II problem.
\end{remark}

\begin{remark}
	The upper and lower bounds for the retrieval rate do not depend explicitly on the transition probabilities of the noisy channels $p(y_n|x_n)$, but rather depend on the capacities of the noisy channels $C_n$.
\end{remark}

\begin{remark}
	Theorem~\ref{Thm1} and Theorem~\ref{Thm2} imply that the channel coding needed for combating channel errors is ``almost seperable" from the retrieval scheme. The channel coding problem and the retrieval problem are coupled only through agreeing on a traffic ratio vector $\bt$. Other than $\bt$, the channel coding acts as an outer code for the responses of the databases to the user queries. Interestingly, the result implies that our schemes work even for heterogeneous channels, e.g., if $N=2$, the channel from one database can be a BSC, and the channel from the other database can be a BEC.  
\end{remark}

\begin{remark}
	Our results imply that randomized strategies for PIR cannot increase the retrieval rate. We can view the noisy channel between the user and the database as a randomizer for the actions of the databases, which is available to the databases but not available to the user. Since the capacity expression does not depend on $p(y_n|x_n)$ and is always maximized by $C_n=1$, any randomizing strategy $p(y_n|x_n)$ cannot enhance the retrieval rate.
\end{remark}

\begin{corollary}[Exact capacity for $M=2$ and $M=3$ messages]\label{M3N2}
	For NPIR, the capacity $C_{\text{PIR}}(\bc)$ for $M=2,3$, and an arbitrary $N$ is given by:
	\begin{align}\label{capacityM32}
	C_{\text{PIR}}(\bc) \!=\!\!  
	\left\{
	\begin{array}{ll}
	\!\!\max_{n_0,n_1 \in \{1, \cdots, N\}} \frac{n_0 n_1}{\sum_{n=1}^{n_0} \frac{n_0+1}{C_n}+\sum_{n=n_0+1}^{n_1} \frac{n_0}{C_n}}, \:\: &M=2 \\
	\!\!\max_{n_0,n_1,n_2 \in \{1, \cdots, N\}} \frac{n_0n_1n_2}{\sum_{n=1}^{n_0} \frac{n_0n_1+n_0+1}{C_n}+\sum_{n=n_0+1}^{n_1} \frac{n_0n_1+n_0}{C_n}+\sum_{n=n_1+1}^{n_2} \frac{n_0n_1}{C_n}}, \:\: &M=3
	\end{array}
	\right.
	\end{align}
\end{corollary}

The proof of Corollary~\ref{M3N2} follows from the optimality of the PIR-WTC-II scheme in \cite{PIR_WTC_II} for $M=2$ and $M=3$ messages by replacing $1-\mu_n$ by $C_n$.

\paragraph{Example: The capacity for NPIR from BSC($p_1$), BSC($p_2$), $N=2$, $M=3$:}
To show how Theorem~\ref{Thm1} reduces to Corollary~\ref{M3N2} for $M=3$, we apply Theorem~\ref{Thm1} to the case of $M=3$, $N=2$, and the links to the user are BSC($p_1$), and BSC($p_2$). From Theorem~\ref{Thm1}, we can write the upper bound for the achievable retrieval rate as:
\begin{align}\label{minmaxM3N2}
R(\mathbf{C})\leq \max_{\bt \in \mathbb{T}} \min_{n_i \in \{1,2\}} \frac{\sum_{n=1}^{N} \tau_n C_n+\frac{\sum_{n=n_1+1}^{N} \tau_n C_n}{n_1}+\frac{\sum_{n=n_2+1}^{N} \tau_n C_n}{n_1n_2}}{1+\frac{1}{n_1}+\frac{1}{n_1n_2}}
\end{align}
where $C_n=1-H(p_n)$.

By observing $\tau_2=1-\tau_1$ and the fact that $C_n$ is monotonically decreasing in $p_n$ for $p_n \in (0,\frac{1}{2})$ (which implies that $p_1 \leq p_2$ satisfies $C_1 \geq C_2$), \eqref{minmaxM3N2} can be explicitly written as the following linear program:
\begin{align}\label{LP32}
\max_{\tau_2,R}  &\quad R \notag\\
\st &\quad R \leq \frac{1}{3}(1-H(p_1))+\left[(1-H(p_2))-\frac{1}{3}(1-H(p_1))\right]\tau_2\notag\\
&\quad R \leq \frac{2}{5}(1-H(p_1))+\left[\frac{4}{5}(1-H(p_2))-\frac{2}{5}(1-H(p_1))\right]\tau_2\notag\\
&\quad R \leq \frac{4}{7}(1-H(p_1))+\left[\frac{4}{7}(1-H(p_2))-\frac{4}{7}(1-H(p_1))\right]\tau_2\notag\\
&\quad 0 \leq \tau_2 \leq 1
\end{align}
The bound corresponding to $n_1=2$, $n_2=1$ is inactive for all values of $(p_1,p_2)$. Since \eqref{LP32} is a linear program, its solution resides at the corner points of the feasible region. The first corner point occurs at $\tau_2^{(1)}=0$, which corresponds to the upper bound $R \leq \frac{1-H(p_1)}{3}$. The second corner point is at the intersection of the first two constraints, i.e., 
\begin{align}
&\frac{1}{3}(1-H(p_1))+\left[(1-H(p_2))-\frac{1}{3}(1-H(p_1))\right]\tau_2^{(2)}\notag\\
&\qquad\qquad\qquad\qquad\qquad\qquad=\frac{2}{5}(1-H(p_1))+\left[\frac{4}{5}(1-H(p_2))-\frac{2}{5}(1-H(p_1))\right]\tau_2^{(2)}
\end{align}
which leads to,
\begin{align}
\tau_2^{(2)}=\frac{1-H(p_1)}{3(1-H(p_2))+(1-H(p_1))}
\end{align} 
which corresponds to the upper bound $R \leq \frac{2}{\frac{3}{1-H(p_1)}+\frac{1}{1-H(p_2)}}$. Similarly, by observing the intersection between the last two constraints, we have the following upper bound  $R \leq \frac{4}{\frac{4}{1-H(p_1)}+\frac{3}{1-H(p_2)}}$, which is achieved at  $\tau_2^{(3)}=\frac{3(1-H(p_1))}{4(1-H(p_2))+3(1-H(p_1))}$. Consequently, an explicit upper bound for the retrieval rate is:
\begin{align}\label{explicit_ub}
R \leq \max\:\left\{\frac{1-H(p_1)}{3}, \frac{2}{\frac{3}{1-H(p_1)}+\frac{1}{1-H(p_2)}}, \frac{4}{\frac{4}{1-H(p_1)}+\frac{3}{1-H(p_2)}}\right\}
\end{align}

\begin{figure}[t]
	\centering
	\includegraphics[width=1\textwidth]{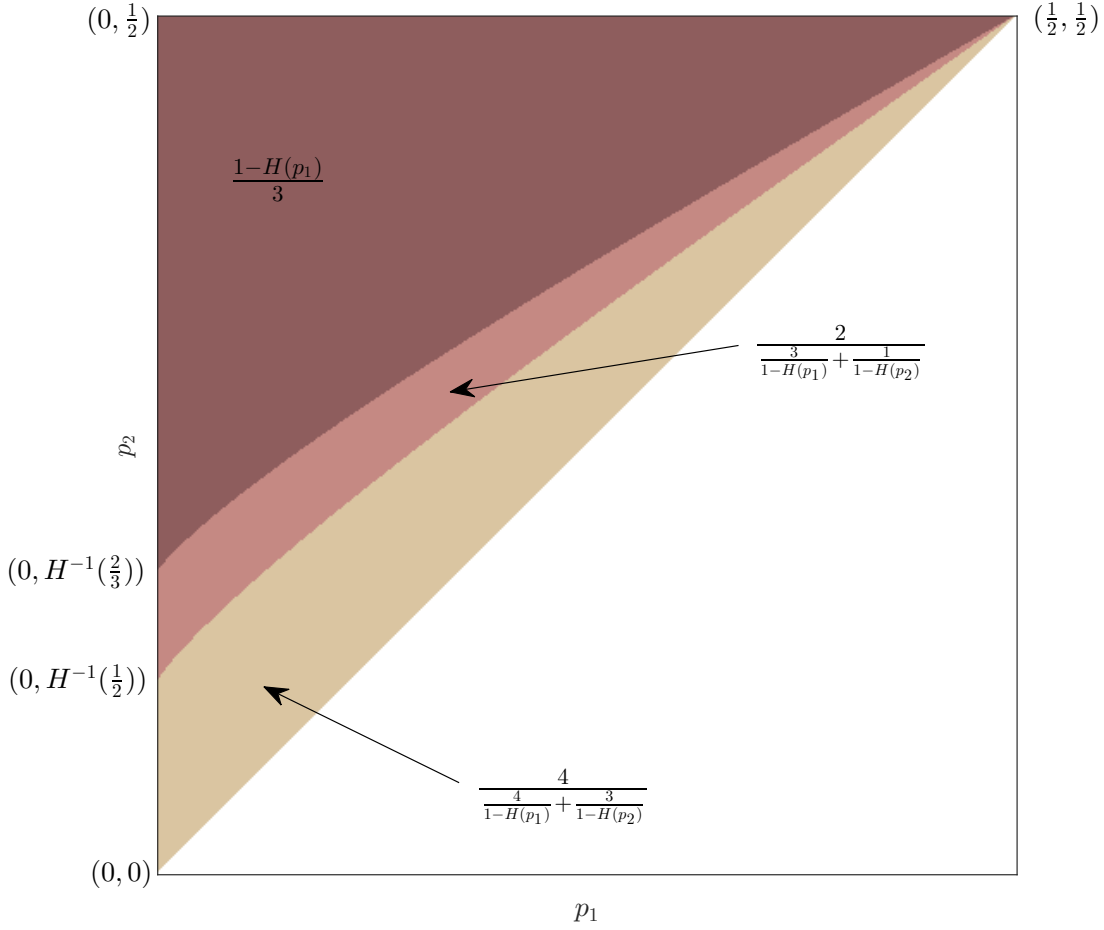}
	\caption{Partitions of $(p_1,p_2)$ space according to retrieval rate expression for $M = 3, N = 2$.}
	\label{regionsM3N2}
	\vspace*{-0.4cm}
\end{figure}

\begin{figure}[t]
	\centering
	\includegraphics[width=1\textwidth]{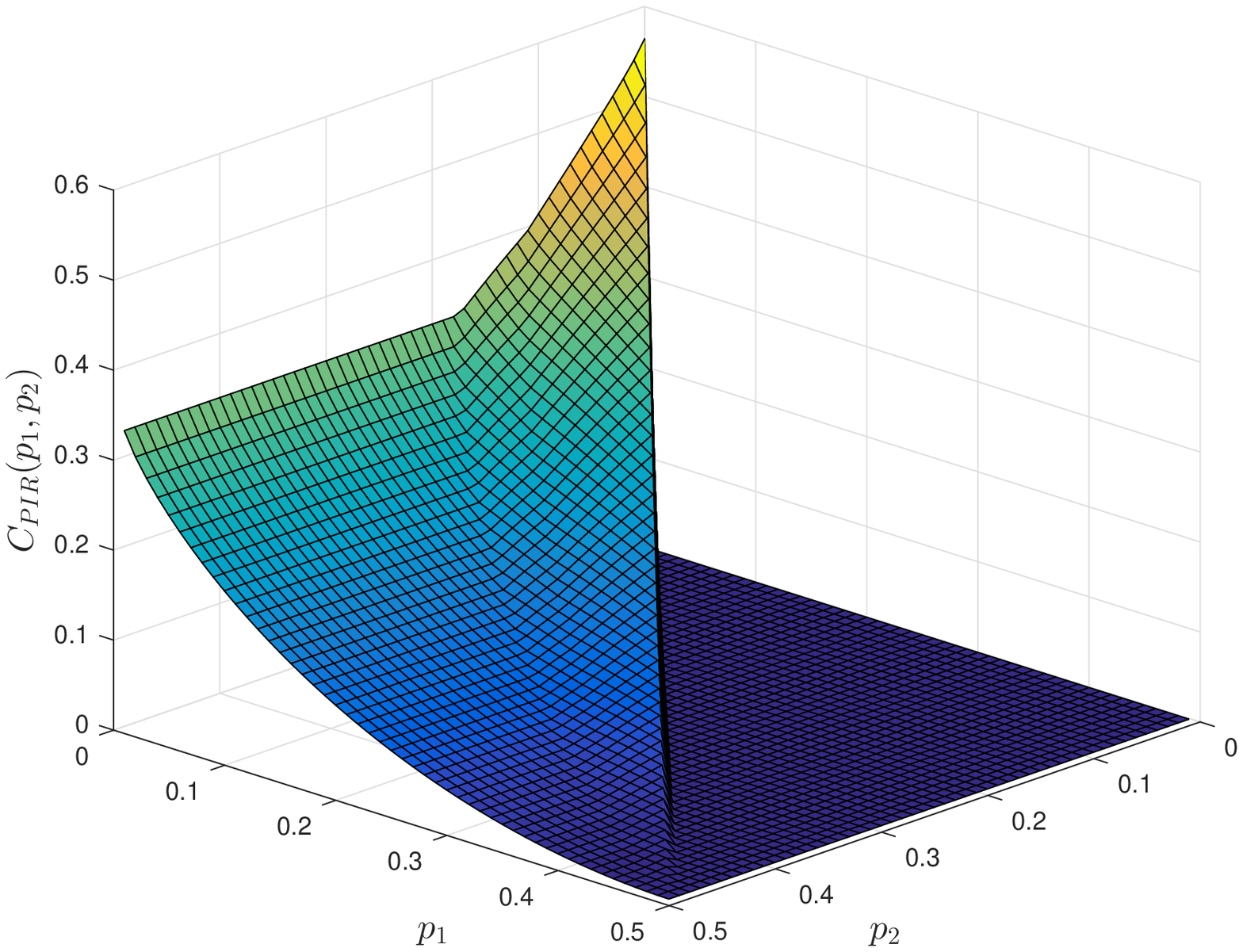}
	\caption{Capacity function $C_{\text{PIR}}(p_1,p_2)$ for $M=3$, $N=2$.}
	\label{capacityM3N2}
	\vspace*{-0.4cm}
\end{figure}

In Section~\ref{BSC}, we will show how these rates can be achieved, hence \eqref{explicit_ub} is the exact capacity. This capacity result is illustrated in Fig.~\ref{regionsM3N2}. The figure shows the partitioning of the $(p_1,p_2)$ (by convention $p_1 \leq p_2$) space according to the active capacity expression. When the ratio $2 <\frac{1-H(p_1)}{1-H(p_2)} \leq 3$, $C_{\text{PIR}}(p_1,p_2)=\frac{2}{\frac{3}{1-H(p_1)}+\frac{1}{1-H(p_2)}}$.  When the ratio $\frac{1-H(p_1)}{1-H(p_2)} \leq 2$, $C_{\text{PIR}}(p_1,p_2)=\frac{4}{\frac{4}{1-H(p_1)}+\frac{3}{1-H(p_2)}}$, otherwise, $C_{\text{PIR}}(p_1,p_2)=\frac{1-H(p_1)}{3}$. Interestingly, Fig.~\ref{regionsM3N2} shows that the dominant strategy for most $(p_1,p_2)$ pairs is to rely only on database~1 for the retrieval process. The capacity function $C_{\text{PIR}}(p_1,p_2)$ is shown in Fig.~\ref{capacityM3N2}. The figure shows that the maximum value for the capacity is $C_{\text{PIR}}(0,0)=\frac{4}{7}$, which is consistent with \cite{JafarPIR}. The figure also shows that $C_{\text{PIR}}(0.5,0.5)=0$, as the answer strings become independent of the user queries. We observe that $C_{\text{PIR}}(0,p_2)=\frac{1}{3}$ for $p_2 \geq H^{-1}(\frac{2}{3})=0.1737$, since the retrieval is performed only from database~1, which is connected to the user via a noiseless link. 

\begin{remark}
	We will show in Section~\ref{achievability} that channel coding and retrieval schemes for NPIR are almost separable. Nevertheless, the final capacity expression couples the capacity of the noisy channels and the retrieval rates from databases with noiseless links in a non-trivial way. We illustrate the capacity expression in \eqref{explicit_ub} by means of circuit theory analogy in Fig.~\ref{fig:circuit noisy}. The current from the current source represents the number of desired bits, the voltage across the current source corresponds to the achievable retrieval rate, and the channel effect of the link connected to the $n$th database is abstracted via a parallel resistor, whose value depends on the capacity of the channel and the total download from the $n$th database. Intuitively, to maximize the retrieval rate, the user chooses one of the three circuits in Fig.~\ref{fig:circuit noisy}. The circuits are arranged ascendingly in the number of the desired bits (namely, 1, 2, 4 bits), while the values of the resistors decrease, as the total download increases and/or due to adding extra parallel branch. This results in a tension between conveying more desired bits and decreasing the equivalent resistor of the circuit. The capacity-achieving scheme is the one which maximizes the product of these contradictory effects (i.e., the voltage). 
	\begin{figure}[h]
		\centering
		\begin{subfigure}[b]{0.32\textwidth}
			\includegraphics[width=\textwidth]{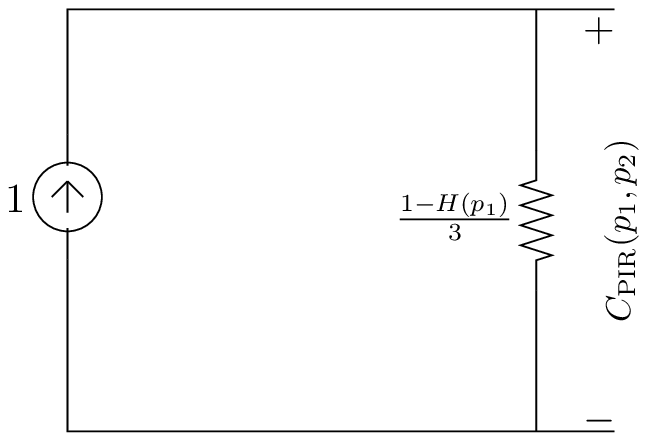}
			\caption{$C_\text{PIR}=\frac{1-H(p_1)}{3}$}
		\end{subfigure}
		\begin{subfigure}[b]{0.32\textwidth}
			\includegraphics[width=\textwidth]{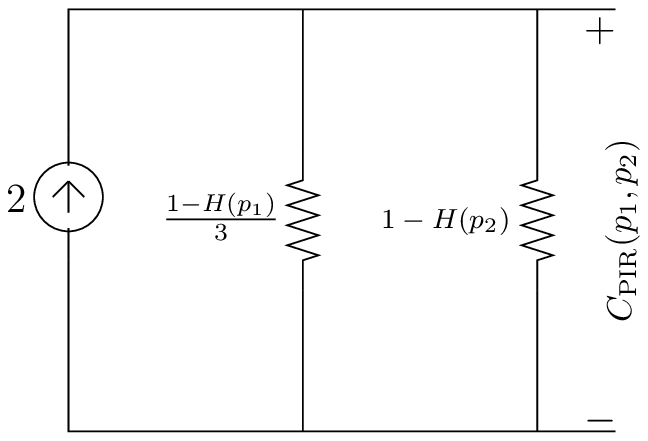}
			\caption{$C_\text{PIR}=\frac{2}{\frac{3}{1-H(p_1)}+\frac{1}{1-H(p_2)}}$}
		\end{subfigure}
		\begin{subfigure}[b]{0.32\textwidth}
			\includegraphics[width=\textwidth]{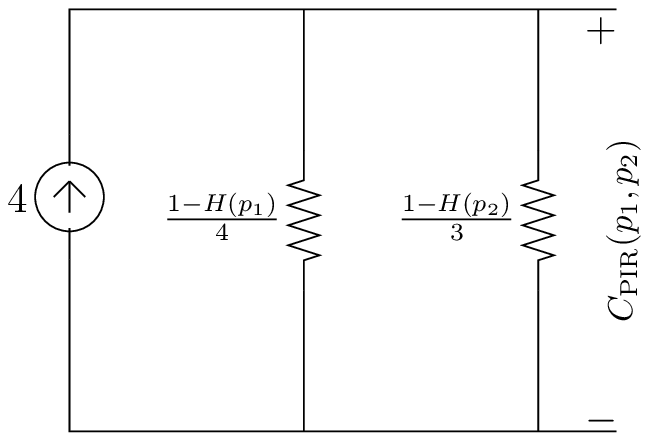}
			\caption{$C_\text{PIR}=\frac{4}{\frac{4}{1-H(p_1)}+\frac{3}{1-H(p_2)}}$}
		\end{subfigure}
		\caption{Circuit analogy for the capacity expression of PIR from BSC($p_1$), BSC($p_2$).}
		\label{fig:circuit noisy}
	\end{figure}
\end{remark}

\section{Converse Proof for NPIR}\label{converse}
In this section, we derive a general upper bound for the NPIR problem. The main idea of the converse hinges on the fact that the traffic from the databases should be dependent on the relative channel qualities (i.e., channel capacities) of the response channels. Thus, we extend the converse proof in \cite{KarimAsymmetricPIR} to account for the noisy observations.

We will need the following lemma, which characterizes the channel effect on the noisy answer strings. The lemma states that the remaining uncertainty on a subset of answer strings after revealing the queries and the message set is a sum of single-letter conditional entropies of the noisy channels over the lengths of the answer strings. The lemma is a consequence of the Markov chain $(W_{1:M},Q_{1:N}^{[m]},\tilde{A}_{1:n-1}^{[m]}) \rightarrow A_n^{[m]} \rightarrow \tilde{A}_n^{[m]}$.  
\begin{lemma}[Channel effect]\label{lemma0}
	For any subset $\mathcal{S} \subseteq \{1, \cdots, N\}$ for all $m \in \{1, \cdots, M\}$, the remaining uncertainty on the noisy answer strings $\tilde{A}_\mathcal{S}^{[m]}$ given $(W_{1:M},Q_{1:N}^{[m]})$ is given by,
	\begin{align}\label{lemma0_result}
	H(\tilde{A}_\mathcal{S}^{[m]}|W_{1:M},Q_{1:N}^{[m]})=\sum_{n \in \mathcal{S}} \sum_{\eta_n=1}^{t_n} H\left(Y_{n,\eta_n}^{[m]}|X_{n,\eta_n}^{[m]}\right)
	\end{align}
	Furthermore, \eqref{lemma0_result} is true if conditioned on the complementary subset of the noisy answer strings $\tilde{A}_{\bar{\mathcal{S}}}^{[m]}$, i.e.,
		\begin{align}\label{lemma0_result1}
		H(\tilde{A}_\mathcal{S}^{[m]}|W_{1:M},Q_{1:N}^{[m]},\tilde{A}_{\bar{\mathcal{S}}}^{[m]})=\sum_{n \in \mathcal{S}} \sum_{\eta_n=1}^{t_n} H\left(Y_{n,\eta_n}^{[m]}|X_{n,\eta_n}^{[m]}\right)
		\end{align}
		where $\bar{\mathcal{S}}=\{1, \cdots, N\}\setminus \mathcal{S}$.
\end{lemma}

\begin{Proof}
	We start with the left hand side of \eqref{lemma0_result},
	\begin{align}
		H(\tilde{A}_\mathcal{S}^{[m]}|W_{1:M},Q_{1:N}^{[m]})=&\sum_{n \in \mathcal{S}} H(\tilde{A}_n^{[m]}|\tilde{A}_{1:n-1}^{[m]},W_{1:M},Q_{1:N}^{[m]}) \\
		\stackrel{\eqref{answer_constraint}}{=}&\sum_{n \in \mathcal{S}}H(\tilde{A}_n^{[m]}|\tilde{A}_{1:n-1}^{[m]},W_{1:M},Q_{1:N}^{[m]},A_n^{[m]}) \label{eq_L0_1}\\
		=&\sum_{n \in \mathcal{S}}H(\tilde{A}_n^{[m]}|A_n^{[m]}) \label{eq_L0_2}\\
		=&\sum_{n \in \mathcal{S}} \sum_{\eta_n=1}^{t_n} H(Y_{n,\eta_n}^{[m]}|X_{n,1}^{[m]}, \cdots, X_{n,t_n}^{[m]},Y_{n,1}, \cdots, Y_{n,\eta_n-1}^{[m]}) \\
		\stackrel{\eqref{memoryless}}{=}& \sum_{n \in \mathcal{S}} \sum_{\eta_n=1}^{t_n} H(Y_{n,\eta_n}^{[m]}|X_{n,\eta_n}^{[m]}) \label{eq_L0_3}
	\end{align}
	where \eqref{eq_L0_1} follows from the fact that $A_n^{[m]}$ is a deterministic function of $(W_{1:M},Q_{n}^{[m]})$, \eqref{eq_L0_2} follows from the fact that $(W_{1:M},Q_{1:N}^{[m]},\tilde{A}_{1:n-1}^{[m]}) \rightarrow A_n^{[m]} \rightarrow \tilde{A}_n^{[m]}$ is a Markov chain, \eqref{eq_L0_3} follows from the fact that the channel is memoryless.
	
	The proof of \eqref{lemma0_result1} follows similarly by observing that $(W_{1:M},Q_{1:N}^{[m]},\tilde{A}_{1:n-1}^{[m]},\tilde{A}_{\bar{\mathcal{S}}}^{[m]}) \rightarrow A_n^{[m]} \rightarrow \tilde{A}_n^{[m]}$ is a Markov chain as well.  
\end{Proof}

We need the following lemma which upper bounds the mutual information between the noisy answer strings  and the interfering messages with a linear function of the channel capacities.
\begin{lemma}[Noisy interference bound]\label{lemma_converse1}
	For NPIR, the mutual information between the interfering messages $W_{2:M}$ and the noisy answer strings $\tilde{A}_{1:N}^{[1]}$ given the desired message $W_1$ is upper bounded by, 	
	\begin{align}
	   I\left(W_{2:M};Q_{1:N}^{[1]}, \tilde{A}_{1:N}^{[1]}|W_{1} \right) \leq \sum_{n=1}^{N} t_nC_n-L+ o(L) \label{eq_L1}
	\end{align}
\end{lemma}

\begin{Proof}
	We start with the left hand side of \eqref{eq_L1},
	\begin{align}
	I(W_{2:M};Q_{1:N}^{[1]},& \tilde{A}_{1:N}^{[1]}|W_{1}) \notag\\
	\stackrel{\eqref{msg_indep}}{=}&I\left(W_{2:M};W_{1},Q_{1:N}^{[1]}, \tilde{A}_{1:N}^{[1]} \right) \label{eq_L1_1}\\
	       =&I\left(W_{2:M};Q_{1:N}^{[1]}, \tilde{A}_{1:N}^{[1]} \right)+I\left(W_{2:M};W_{1}|Q_{1:N}^{[1]}, \tilde{A}_{1:N}^{[1]} \right)\\
	       \stackrel{\eqref{reliability_constraint}}{\leq}&I\left(W_{2:M};Q_{1:N}^{[1]}, \tilde{A}_{1:N}^{[1]} \right)+o(L) \label{eq_L1_3}\\
	       \stackrel{\eqref{independency}}{=}&I\left(W_{2:M}; \tilde{A}_{1:N}^{[1]} |Q_{1:N}^{[1]}\right)+o(L) \label{eq_L1_2}\\
	       =&H\left(\tilde{A}_{1:N}^{[1]}|Q_{1:N}^{[1]}\right)-H\left(\tilde{A}_{1:N}^{[1]}|W_{2:M},Q_{1:N}^{[1]}\right)+o(L)\\
	       =&H\left(\tilde{A}_{1:N}^{[1]}|Q_{1:N}^{[1]}\right)-H\left(\tilde{A}_{1:N}^{[1]},W_1|W_{2:M},Q_{1:N}^{[1]}\right)\notag\\
	       &+H\left(W_1|W_{2:M},Q_{1:N}^{[1]},\tilde{A}_{1:N}^{[1]}\right)+o(L)\\
	       \stackrel{\eqref{reliability_constraint}}{\leq}& H\left(\tilde{A}_{1:N}^{[1]}|Q_{1:N}^{[1]}\right)-H\left(\tilde{A}_{1:N}^{[1]},W_1|W_{2:M},Q_{1:N}^{[1]}\right)+o(L)\label{eq_L1_4}\\
	       =&H\left(\tilde{A}_{1:N}^{[1]}|Q_{1:N}^{[1]}\right)-H\left(W_1|W_{2:M},Q_{1:N}^{[1]}\right)-H\left(\tilde{A}_{1:N}^{[1]}|W_{1:M},Q_{1:N}^{[1]}\right)+o(L)\\
	       \stackrel{\eqref{lemma0_result}}{\leq}&\sum_{n=1}^N \sum_{\eta_n=1}^{t_n} \left[H\left(Y_{n,\eta_n}^{[1]}\right)-H\left(Y_{n,\eta_n}^{[1]}|X_{n,\eta_n}^{[1]}\right)\right]-L+o(L) \label{eq_L1_5}\\
	       =&\sum_{n=1}^N \sum_{\eta_n=1}^{t_n} I\left(X_{n,\eta_n}^{[1]};Y_{n,\eta_n}^{[1]}\right)-L+o(L) \\
	       \leq & \sum_{n=1}^N t_n C_n -L+o(L) \label{eq_L1_6}
	\end{align}
	where \eqref{eq_L1_1} follows from the independence of the messages, \eqref{eq_L1_3}, \eqref{eq_L1_4} follow from the decodability of $W_1$ given $(Q_{1:N}^{[1]},\tilde{A}_{1:N}^{[1]})$, \eqref{eq_L1_2} follows from the independence of $(W_{2:M},Q_{1:N}^{[1]})$, \eqref{eq_L1_5} follows from the independence of $(W_1,W_{2:M},Q_{1:N}^{[1]})$, Lemma~\ref{lemma0}, and the fact that conditioning cannot increase entropy, \eqref{eq_L1_6} follows from the fact that $I\left(X_{n,\eta_n}^{[m]};Y_{n,\eta_n}^{[m]}\right) \leq C_n$ by the definition of the $n$th channel capacity.
\end{Proof}

Finally, in order to capture the recursive structure of the problem in terms of the messages and to express the potential asymmetry of the optimal scheme, we will need the following lemma, which inductively lower bounds the mutual information term in Lemma~\ref{lemma_converse1}. The lemma implies that $n_{m-1}$ databases can apply a symmetric scheme when the retrieval problem is reduced to retrieving message $W_{m-1}$ from the set of $W_{m-1:M}$ messages. For the remaining answer strings, we directly bound them by their corresponding length of the unobserved portion $\sum_{n=n_{m-1}+1}^N t_n C_n$.    
\begin{lemma}[Noisy induction lemma]\label{lemma_converse2}
	For all $m\in \{2,\dots,M\}$ and for an arbitrary $n_{m-1} \in \{1, \cdots, N\}$, the mutual information term in Lemma~\ref{lemma_converse1} can be inductively lower bounded as,
	\begin{align} \label{eq_L2}
	&I\left( W_{m:M} ; Q_{1:N}^{[m-1]}, \tilde{A}_{1:N}^{[m-1]} | W_{1:m-1} \right)  \notag \\
	&\quad\quad\quad \geq \frac{1}{n_{m-1}}  I\left(W_{m+1:M}; Q_{1:N}^{[m]}, \tilde{A}_{1:N}^{[m]}|W_{1:m} \right)  +\frac{1}{n_{m-1}} \left(L-\!\!\!\!\!\!\sum_{n=n_{m-1}+1}^N \!\!\!\!\!\!t_n C_n \right)-\frac{o(L)}{n_{m-1}}
	\end{align}
\end{lemma}

\begin{Proof}
	We start with the left hand side of \eqref{eq_L2} after multiplying by $n_{m-1}$,
	\begin{align}
	& n_{m-1}\,I\left(W_{m:M} ; Q_{1:N}^{[m-1]}, \tilde{A}_{1:N}^{[m-1]}| W_{1:m-1}  \right) \notag \\
	&\qquad \geq  n_{m-1}\, I\left(W_{m:M} ; Q_{1:n_{m-1}}^{[m-1]}, \tilde{A}_{1:n_{m-1}}^{[m-1]}| W_{1:m-1}  \right) \label{eq_IL_12}\\
	& \label{eq_IL_1}\qquad \geq  \sum_{n=1}^{n_{m-1}} I\left(W_{m:M} ; Q_n^{[m-1]}, \tilde{A}_n^{[m-1]}| W_{1:m-1}  \right) \\
	& \label{eq_IL_2}\qquad \stackrel{\eqref{privacy_constraint}}{=}  \sum_{n=1}^{n_{m-1}} I\left(W_{m:M} ; Q_n^{[m]}, \tilde{A}_n^{[m]}| W_{1:m-1}  \right) \\
	& \label{eq_IL_444}\qquad \stackrel{\eqref{independency}}{=}  \sum_{n=1}^{n_{m-1}} I\left(W_{m:M} ;  \tilde{A}_n^{[m]}|Q_n^{[m]}, W_{1:m-1}  \right)\\
	& \label{eq_IL_44}\qquad =  \sum_{n=1}^{n_{m-1}} H\left(\tilde{A}_n^{[m]}|Q_n^{[m]}, W_{1:m-1}  \right) -H\left(\tilde{A}_n^{[m]}|Q_n^{[m]}, W_{1:M}  \right)\\
	& \label{eq_IL_3}\qquad \geq \sum_{n=1}^{n_{m-1}} H\left(\tilde{A}_n^{[m]}|\tilde{A}^{[m]}_{1:n-1},Q_{1:n_{m-1}}^{[m]}, W_{1:m-1}  \right)-H\left(\tilde{A}_n^{[m]}|\tilde{A}^{[m]}_{1:n-1},Q_{1:n_{m-1}}^{[m]}, W_{1:M}  \right) \\
	& \label{eq_IL_4}\qquad =  \sum_{n=1}^{n_{m-1}} I\left(W_{m:M};\tilde{A}_n^{[m]}|\tilde{A}^{[m]}_{1:n-1},Q_{1:n_{m-1}}^{[m]}, W_{1:m-1}  \right) \\
	&\label{eq_IL_5}  \qquad = I\left(W_{m:M}; \tilde{A}_{1:n_{m-1}}^{[m]} | Q_{1:n_{m-1}}^{[m]}, W_{1:m-1}  \right) \\
	&\label{eq_IL_6} \qquad \stackrel{\eqref{independency}}{=} I\left(W_{m:M}; Q_{1:n_{m-1}}^{[m]}, \tilde{A}_{1:n_{m-1}}^{[m]} |  W_{1:m-1}  \right)\\
	&\label{eq_IL_65}\quad\:\: \stackrel{\eqref{independency},\eqref{answer_constraint}}{=}I\left(W_{m:M}; Q_{1:N}^{[m]}, \tilde{A}_{1:N}^{[m]} |  W_{1:m-1}\right)\!-\!I\left(W_{m:M}; \tilde{A}_{n_{m-1}+1:N}^{[m]} | Q_{1:N}^{[m]}, \tilde{A}_{1:n_{m-1}}^{[m]}, W_{1:m-1}  \right)\\
	&\qquad \label{eq_IL_66} =I\left(W_{m:M}; Q_{1:N}^{[m]}, \tilde{A}_{1:N}^{[m]} |  W_{1:m-1}\right)-H\left(\tilde{A}_{n_{m-1}+1:N}^{[m]} | Q_{1:N}^{[m]}, \tilde{A}_{1:n_{m-1}}^{[m]}, W_{1:m-1}  \right)\notag\\
	&\qquad\quad+H\left(\tilde{A}_{n_{m-1}+1:N}^{[m]} | Q_{1:N}^{[m]}, \tilde{A}_{1:n_{m-1}}^{[m]}, W_{1:M}  \right)\\
	&\qquad \label{eq_IL_67}\stackrel{\eqref{lemma0_result1}}{\geq}I\left(W_{m:M}; Q_{1:N}^{[m]}, \tilde{A}_{1:N}^{[m]} |  W_{1:m-1}\right)-\sum_{n=n_{m-1}+1}^N \sum_{\eta_n=1}^{t_n} \left[H\left(Y_{n,\eta_n}^{[m]}\right)-H\left(Y_{n,\eta_n}^{[m]}|X_{n,\eta_n}^{[m]}\right)\right]\\
	& \label{eq_IL_7} \qquad \stackrel{\eqref{reliability_constraint}}{\geq} I\left(W_{m:M}; W_m, Q_{1:N}^{[m]}, \tilde{A}_{1:N}^{[m]} |  W_{1:m-1}\right)-\sum_{n=n_{m-1}+1}^N \sum_{\eta_n=1}^{t_n} I\left(X_{n,\eta_n}^{[m]};Y_{n,\eta_n}^{[m]}\right)-o(L)\\
	& \qquad =I\left(W_{m:M};W_m|  W_{1:m-1}\right)+I\left(W_{m:M};Q_{1:N}^{[m]}, \tilde{A}_{1:N}^{[m]} |  W_{1:m}\right)\notag\\
	&\qquad\quad-\sum_{n=n_{m-1}+1}^N \sum_{\eta_n=1}^{t_n} I\left(X_{n,\eta_n}^{[m]};Y_{n,\eta_n}^{[m]}\right) -o(L)\\
	& \qquad =I\left(W_{m+1:M}; Q_{1:N}^{[m]}, \tilde{A}_{1:N}^{[m]}|W_{1:m} \right)  + \left(L-\sum_{n=n_{m-1}+1}^N \sum_{\eta_n=1}^{t_n} I\left(X_{n,\eta_n}^{[m]};Y_{n,\eta_n}^{[m]}\right) \right)-o(L)\\
	&\qquad \geq I\left(W_{m+1:M}; Q_{1:N}^{[m]}, \tilde{A}_{1:N}^{[m]}|W_{1:m} \right)  + \left(L-\sum_{n=n_{m-1}+1}^N t_n C_n\right)-o(L) \label{eq_IL_8}
	\end{align}
where \eqref{eq_IL_12}, \eqref{eq_IL_1} follow from the non-negativity of mutual information, \eqref{eq_IL_2} follows from the privacy constraint, \eqref{eq_IL_444} follows from the independence of $\left(W_{m:M},Q_n^{[m]}\right)$, \eqref{eq_IL_3} follows from the fact that conditioning cannot increase entropy and from the fact that $(W_{1:M},Q_{1:n_{m-1}}^{[m]},\tilde{A}_{1:n-1}^{[m]}) \rightarrow (W_{1:M},Q_n^{[m]}) \rightarrow \tilde{A}_n^{[m]}$ forms a Markov chain, \eqref{eq_IL_6} follows from the independence of the messages and the queries, \eqref{eq_IL_65} follows from the chain rule, the independence of the queries and the messages, and the fact that $Q_{1:N}^{[m]} \rightarrow Q_{1:n_{m-1}}^{[m]} \rightarrow \tilde{A}_{1:n_{m-1}}^{[m]}$ forms a Markov chain by (\ref{answer_constraint}), \eqref{eq_IL_67} follows from the fact that conditioning reduces entropy and Lemma~\ref{lemma0}, \eqref{eq_IL_7} follows from the reliability constraint, \eqref{eq_IL_8} follows from the definition of the channel capacity. Finally, dividing both sides by $n_{m-1}$ leads to \eqref{eq_L2}.
\end{Proof}

Now, we are ready to derive an explicit upper bound for the retrieval rate from noisy channels. Fixing the length of the $n$th answer string to $t_n$ and applying Lemma~\ref{lemma_converse1} and Lemma~\ref{lemma_converse2} successively for an arbitrary sequence $\{n_i\}_{i=1}^{M-1} \subset \{1, \cdots, N\}^{M-1}$, we have the following,
\begin{align}
&\sum_{n=1}^{N} t_nC_n-L+\tilde{o}(L)  \notag \\
&\label{eq_induction1}\quad \stackrel{\eqref{eq_L1}}{\geq} I\left(W_{2:M}; Q_{1:N}^{[1]}, \tilde{A}_{1:N}^{[1]}|W_{1} \right) \\
&\quad \stackrel{\eqref{eq_L2}}{\geq}\frac{1}{n_{1}} \left(\!L\!-\!\!\!\sum_{n=n_{1}+1}^N \!\!\! t_nC_n \right)+ \frac{1}{n_{1}}  I\left(W_{3:M}; Q_{1:N}^{[2]}, \tilde{A}_{1:N}^{[2]}|W_{1:2} \right)    \\
&\quad \stackrel{\eqref{eq_L2}}{\geq}\frac{1}{n_{1}} \left(\!L\!-\!\!\!\sum_{n=n_{1}+1}^N \!\!\! t_nC_n \right)+\frac{1}{n_{1}n_2}\left(\!L\!-\!\!\!\sum_{n=n_{2}+1}^N \!\!\! t_nC_n \right)+ \frac{1}{n_{2}}  I\left(W_{4:M}; Q_{1:N}^{[3]}, \tilde{A}_{1:N}^{[3]}|W_{1:3} \right)    \\
&\quad \stackrel{\eqref{eq_L2}}{\geq} \dots \notag\\
&\quad \stackrel{\eqref{eq_L2}}{\geq}\frac{1}{n_{1}} \left(\!L\!-\!\!\!\sum_{n=n_{1}+1}^N \!\!\! t_nC_n \right)+\frac{1}{n_{1}n_2}\left(\!L\!-\!\!\!\sum_{n=n_{2}+1}^N \!\!\! t_nC_n \right)\!+\!  \cdots \!+\!\frac{1}{\prod_{i=1}^{M-1} n_i}\!\left(\!L\!-\!\!\!\!\!\sum_{n=n_{M-1}+1}^N \!\!\!\!\! t_nC_n \right)
\end{align}
where $\tilde{o}(L)=\left(1+\frac{1}{n_1}+\frac{1}{n_1n_2}+\cdots+\frac{1}{\prod_{i=1}^{M-1}n_i}\right)o(L)$, \eqref{eq_induction1} follows from Lemma~\ref{lemma_converse1}, and the remaining bounding steps follow from successive application of Lemma~\ref{lemma_converse2}.

Ordering terms, we have,
\begin{align}
\left(1+\frac{1}{n_1}\!+\!\frac{1}{n_1n_2}\!+\!\cdots\!+\!\frac{1}{\prod_{i=1}^{M-1}n_i}\right)\!L \leq
\left(\!\theta(0)\!+\!\frac{\theta(n_1)}{n_1}+\!\cdots\!+\frac{\theta(n_{M-1})}{\prod_{i=1}^{M-1} n_i}\right)\!\sum_{n=1}^N t_n \!+\! \tilde{o}(L)
\end{align}
where $\theta(\ell)=\sum_{n=\ell+1}^{N} \tau_nC_n$ 

We conclude the proof by taking $L \rightarrow \infty$. Thus, for an arbitrary sequence $\{n_i\}_{i=1}^{M-1}$, we have
\begin{align}
R(\boldsymbol{\tau}, \mathbf{C})&= \frac{L}{ \sum_{n=1}^{N} t_n } \leq  \frac{\theta(0)+\frac{\theta(n_1)}{n_1}+\frac{\theta(n_2)}{n_1n_2}+\cdots+\frac{\theta(n_{M-1})}{\prod_{i=1}^{M-1} n_i}}{1+\frac{1}{n_1}+\frac{1}{n_1n_2}+\cdots+\frac{1}{\prod_{i=1}^{M-1}n_i}}
\end{align}
Finally, we get the tightest bound by minimizing over the sequence $\{n_i\}_{i=1}^{M-1}$ over the set $\{1, \cdots, N\}$, as

\begin{align}
R(\boldsymbol{\tau}, \mathbf{C}) &\leq \min_{n_i \in \{1, \cdots, N\}} \frac{\theta(0)+\frac{\theta(n_1)}{n_1}+\frac{\theta(n_2)}{n_1n_2}+\cdots+\frac{\theta(n_{M-1})}{\prod_{i=1}^{M-1} n_i}}{1+\frac{1}{n_1}+\frac{1}{n_1n_2}+\cdots+\frac{1}{\prod_{i=1}^{M-1}n_i}}\\
   &=\min_{n_i \in \{1, \cdots, N\}} \frac{\sum_{n=1}^{N} \tau_nC_n+\frac{\sum_{n=n_1+1}^{N} \tau_n C_n}{n_1}+\frac{\sum_{n=n_2+1}^{N} \tau_n C_n}{n_1n_2}+\cdots+\frac{\sum_{n=n_{M-1}+1}^{N} \tau_n C_n}{\prod_{i=1}^{M-1} n_i}}{1+\frac{1}{n_1}+\frac{1}{n_1n_2}+\cdots+\frac{1}{\prod_{i=1}^{M-1}n_i}}
\end{align}

The user and the databases can agree on a traffic ratio vector $\bt \in \mathbb{T}=\{(\tau_1, \cdots, \tau_N): \tau_n \geq 0 \:\: \forall n, \sum_{n=1}^{N} \tau_n=1\}$ that maximizes $R(\boldsymbol{\tau}, \mathbf{C})$, hence the retrieval rate $R(\mathbf{C})$ is upper bounded by,

\begin{align}
R(\mathbf{C}) &\leq \max_{\bt \in \mathbb{T}}\quad R(\boldsymbol{\tau}, \mathbf{C}) \\ 
              &=\max_{\bt \in \mathbb{T}} \min_{n_i \in \{1, \cdots, N\}} \frac{\sum_{n=1}^{N} \tau_nC_n+\frac{\sum_{n=n_1+1}^{N} \tau_n C_n}{n_1}+\frac{\sum_{n=n_2+1}^{N} \tau_n C_n}{n_1n_2}+\cdots+\frac{\sum_{n=n_{M-1}+1}^{N} \tau_n C_n}{\prod_{i=1}^{M-1} n_i}}{1+\frac{1}{n_1}+\frac{1}{n_1n_2}+\cdots+\frac{1}{\prod_{i=1}^{M-1}n_i}}
\end{align}

\section{Achievability Proof for NPIR}\label{achievability}
In this section, we present the achievability proof for the NPIR problem. We show that by means of the random coding argument, each database can independently encode its response such that the probability of error can be made vanishingly small. The databases use the uncoded responses as an indexing mechanism for choosing codewords from a randomly generated codebook. The uncoded responses, which are the truthful responses to the user queries, vary in length to maximize the retrieval rate. The query structure builds on the achievability proofs for PIR under asymmetric traffic constraints \cite{KarimAsymmetricPIR}.  
\subsection{Motivating Example: $M=3$, $N=2$, via BSC($p_1$), BSC($p_2$)}\label{BSC}
We illustrate the retrieval scheme for $N=2$ databases, $M=3$ messages when the answer strings pass through BSC($p_1$) and BSC($p_2$). We show that the channel coding (using linear block codes) is \emph{almost separable} from the retrieval scheme (which hinges on the result of \cite{KarimAsymmetricPIR}). We begin with the case when $(p_1,p_2)=(0.1,0.2)$, then we extend this technique for all $(p_1,p_2)$ pairs. We will need the following lemma, which shows the achievability of Shannon's channel coding theorem for BSC using linear block codes \cite[Theorem~4.17, Corollary~4.18]{RothCoding}.  

\begin{lemma}[Shannon's coding theorem for BSC \cite{RothCoding}]\label{lemma_BSC}
	For BSC($p$) with crossover probability $p \in (0, \frac{1}{2})$. Let $n$, $k$ be integers such that $R=\frac{k}{n} <1-H(p)$, and let $\mathbb{E}_\mathcal{C}[P_e(\mathcal{C})]$ denote the expected probability of error $P_e(\mathcal{C})$ calculated over all linear $[n,k]$ codes $\mathcal{C}$, assuming a nearest-codeword decoder. Then,
	\begin{align}
	\mathbb{E}_\mathcal{C}[P_e(\mathcal{C})] < 2 \cdot 2^{-n\Delta(p,R)}
	\end{align}
	for some $\Delta(p,R)>0$. Moreover, for all $\rho \in (0,1]$, all but less than $\rho$ of the linear $[n,k]$ codes satisfy,
	\begin{align}\label{Pe}
	P_e(\mathcal{C}) < \frac{2}{\rho} \cdot 2^{-n\Delta(p,R)}
	\end{align}  
\end{lemma}

The result implies that as long as the rate of the linear $[n,k]$ code is strictly less than the capacity, then there exists a linear $[n,k]$ code with exponentially decreasing probability of error in $n$ with high probability.

\subsubsection{Achievable Scheme for BSC(0.1), BSC(0.2)}\label{BSC(0.1,0.2)}
Now, we focus on the case when $(p_1,p_2)=(0.1,0.2)$. Using the explicit upper bound in \eqref{explicit_ub}, we infer that $R\leq \frac{4}{\frac{4}{1-H(p_1)}+\frac{3}{1-H(p_2)}}$ which is $0.2183$ for $p_1=0.1$, $p_2=0.2$. To operate at $\tau_2=\tau_2^{(3)}=\frac{3(1-H(p_1))}{4(1-H(p_2))+3(1-H(p_1))}$, we enforce the ratio between the uncoded traffic, i.e., before channel coding, to be $4:3$. This results in coded traffic ratio of $\frac{4}{1-H(p_1)}:\frac{3}{1-H(p_2)}$, which appears in the denominator of the upper bound. Concurrently, this results in retrieving 4 desired bits per scheme repetition, which appears in the numerator.

To that end, the user repeats the following retrieval scheme for $\nu$ times. Each repetition of the scheme operates over blocks of $L^*=4$ bits from all messages $W_{1:3}$. The user permutes the indices of the bits of each message independently and uniformly.  Let $a_i(j)$, $b_i(j)$, $c_i(j)$ denote the $i$th bit of block $j$ from the permuted message $W_1$, $W_2$, $W_3$, respectively. Assume without loss of generality that the desired file is $W_1$. In block $j$, the user requests to download a single bit from each message from database~1, i.e., the user requests to download $a_1(j)$, $b_1(j)$, and $c_1(j)$ from database~1. From database~2, the user exploits the side information generated from database~1 by requesting to download the sums $a_2(j)+b_1(j)$, $a_3(j)+c_1(j)$, and $b_2(j)+c_2(j)$. Finally, the user exploits the side information generated from database~1 by downloading $a_4(j)+b_2(j)+c_2(j)$ from database~2. The query table for the $j$th block is summarized in Table~\ref{tableM3N2_43}. Denote the number of uncoded bits requested from the $n$th database by $D_n$, then $D_1=4$, $D_2=3$. This guarantees that the ratio between the uncoded traffic is $4:3$ (for any number of repetitions $\nu$). This query structure is private, as all combinations of the sums are included in the queries and the indices of the message bits are uniformly and independently permuted for each block of messages (which operate on different set of bits), the privacy constraint is satisfied. 
\begin{table}[h]
	\centering
	\caption{The query table for the $j$th block of $M=3$, $N=2$, $p_1=0.1$, $p_2=0.2$.}
	\label{tableM3N2_43}
	\begin{tabular}{|c|c|}
		\hline
		Database 1 & Database 2 \\
		\hline
		$a_1(j)$  & $a_2(j)+b_1(j)$\\
		$b_1(j)$  & $a_3(j)+c_1(j)$\\
		$c_1(j)$  & $b_2(j)+c_2(j)$\\
		\hline
		$a_4(j)+b_2(j)+c_2(j)$ &       \\
		\hline
	\end{tabular}
\end{table}

After receiving the queries of the user, the $n$th database concatenates the uncoded binary answer strings into a vector $U_n^{[1]}$ of length $\nu D_n$, i.e.,
\begin{align}
U_1^{[1]}=&[a_1(1) \quad b_1(1) \quad c_1(1) \quad a_4(1)+b_2(1)+c_2(1) \notag\\
 &\quad  \cdots \quad  a_1(\nu) \quad b_1(\nu) \quad c_1(\nu) \quad a_4(\nu)+b_2(\nu)+c_2(\nu)]^T \\
U_2^{[1]}=&[a_2(1)+b_1(1) \quad a_3(1)+c_1(1) \quad b_2(1)+c_2(1) \notag\\
& \quad \cdots \quad a_2(\nu)+b_1(\nu) \quad a_3(\nu)+c_1(\nu) \quad b_2(\nu)+c_2(\nu)]^T
\end{align}
The $n$th database encodes the vector $U_n^{[1]}$ to a coded answer string $A_n^{[1]}$ of length $t_n$ using a $(t_n,\nu D_n)$ linear block code (which belongs to the set of good codes that satisfy \eqref{Pe}) such that:
\begin{align}
t_n=\left\lceil \frac{\nu D_n}{1-H(p_n)}\right\rceil
\end{align}
This ensures that $\frac{\nu D_n}{t_n} < 1-H(p_n)$. The $n$th database responds with $A_n^{[1]}$ via the noisy channel BSC($p_n$). The user receives the noisy answer string $\tilde{A}_n^{[1]}$ from the $n$th database.

To perform the decoding, the user employs the nearest-codeword decoder to find an estimate of $A_n^{[1]}$ based on the observation $\tilde{A}_n^{[1]}$. Since $\frac{\nu D_n}{t_n} < 1-H(p_n)$, using Lemma~\ref{lemma_BSC} and the union bound, the probability of error in decoding is upper bounded by:
\begin{align}
P_e(L) &\leq P_e(\mathcal{C}_1)+P_e(\mathcal{C}_2) \\
       &\leq \frac{2}{\rho} \left[2^{-t_1\Delta\left(p_1,\frac{\nu D_1}{t_1}\right)}+2^{-t_2\Delta\left(p_2,\frac{\nu D_2}{t_2}\right)}\right]
\end{align} 

As $\nu \rightarrow \infty$, $L \rightarrow \infty$ and $t_n \rightarrow \infty$, we have $P_e(L) \rightarrow 0$. This ensures the decodability of $U_n^{[1]}$ with high probability. Since the vectors $U_1^{[1]}$, $U_2^{[2]}$ are designed to exploit the side information, the user can cancel the effect of the undesired messages and be left only with the correct $W_1$ with probability of error $P_e(L)$. This satisfies the reliability constraint.

Finally, we calculate the achievable retrieval rate. The retrieval scheme decodes $L=\nu L^*=4\nu$ bits from the desired messages. The retrieval scheme downloads $t_n=\left\lceil \frac{\nu D_n}{1-H(p_n)}\right\rceil$ bits from the $n$th database, hence as $\nu \rightarrow \infty$, we have
\begin{align}
R&= \frac{L}{t_1+t_2}\\
 &= \frac{\nu L^*}{\frac{\nu D_1}{1-H(p_1)}+\frac{\nu D_2}{1-H(p_2)}} \\
 &= \frac{4}{\frac{4}{1-H(p_1)}+\frac{3}{1-H(p_2)}}=0.2183
\end{align}
which matches the upper bound.

\subsubsection{Achieving the Upper Bound for Arbitrary $(p_1,p_2)$}
Now, we show that the upper bound in \eqref{explicit_ub} is achievable for any $(p_1,p_2)$. The idea is to design the uncoded response vectors $U_1^{[1]}$, $U_2^{[2]}$ such that the ratio of their traffic matches one of the corner points of the PIR problem under asymmetric traffic constraints \cite{KarimAsymmetricPIR}.

\paragraph{For $R=\frac{1-H(p_1)}{3}$:} For this rate, the user requests to download from database~1 only and does not access database~2. Thus, the user downloads all the contents of database~1 to satisfy the privacy constraint. Specifically, the user downloads $a_1(j), b_1(j), c_1(j)$ at the $j$th block of the retrieval process. Database~1 encodes the responses $U_1^{[1]}$ into $t_1$-length answer string using $(t_1,\nu D_1)$, where $D_1=3$, and $t_1=\left\lceil \frac{\nu D_1}{1-H(p_1)}\right\rceil$. The user decodes $\nu$ desired symbols from $\nu$ repetitions with vanishingly small probability of error. Consequently, $R=\frac{1-H(p_1)}{3}$.

\paragraph{For $R=\frac{2}{\frac{3}{1-H(p_1)}+\frac{1}{1-H(p_2)}}$:} For this rate, the user designs the queries such that the traffic ratio between the uncoded responses is $3:1$. Thus, in the $j$th block, the user requests to download one bit from each message, i.e., the user requests to download $a_1(j), b_1(j), c_1(j)$ from database~1. The user mixes the undesired information obtained from database~1 into one combined symbol $b_1(j)+c_1(j)$ and uses this symbol as a side information in database~2 by requesting to download $a_2(j)+b_1(j)+c_1(j)$. The query table for the $j$th block of the scheme is depicted in Table~\ref{tableM3N2_31}.  

\begin{table}[h]
	\centering
	\caption{The query table for the $j$th block of $M=3$, $N=2$ to achieve $R=\frac{2}{\frac{3}{1-H(p_1)}+\frac{1}{1-H(p_2)}}$}
	\label{tableM3N2_31}
	\begin{tabular}{|c|c|}
		\hline
		Database 1 & Database 2 \\
		\hline
		$a_1(j),b_1(j),c_1(j)$ & $a_2(j)+b_1(j)+c_1(j)$\\
		\hline
	\end{tabular}
\end{table}

After repeating the retrieval process $\nu$ times, database~1 encodes the responses using a linear $(t_1,\nu D_1)=\left(\left\lceil \frac{3\nu}{1-H(p_1)}\right\rceil,3\nu\right)$ code, while database~2 encodes its responses using a linear $(t_2,\nu D_2)=\left(\left\lceil \frac{\nu}{1-H(p_2)}\right\rceil,\nu\right)$ code. Using Lemma~4, the user can decode the correct $W_1$ with vanishingly small probability of error. The user decodes $L=2\nu$ bits from $W_1$, hence, as $\nu \rightarrow \infty$
\begin{align}
R=\frac{L}{t_1+t_2}=\frac{2}{\frac{3}{1-H(p_1)}+\frac{1}{1-H(p_2)}}
\end{align}

\paragraph{For $R=\frac{4}{\frac{4}{1-H(p_1)}+\frac{3}{1-H(p_2)}}$:} An instance for this scheme is the $(p_1,p_2)=(0.1,0.2)$ example. Please refer to Section~\ref{BSC(0.1,0.2)} for the details.

Therefore, the capacity of the PIR problem from BSC($p_1$), BSC($p_2$) is given by:
\begin{align}
C_{\text{PIR}}(p_1,p_2) = \max\:\left\{\frac{1-H(p_1)}{3}, \frac{2}{\frac{3}{1-H(p_1)}+\frac{1}{1-H(p_2)}}, \frac{4}{\frac{4}{1-H(p_1)}+\frac{3}{1-H(p_2)}}\right\}
\end{align}

\subsection{General Achievable Scheme}
In this section, we present a general achievable scheme for the NPIR problem. The main idea of the scheme is to use the uncoded response from the $n$th database to user's query as an \emph{index} for choosing the transmitted codeword from a codebook generated according to the optimal probability distribution. The query structure maps to one of the corner points of PIR under asymmetric traffic constraints \cite{KarimAsymmetricPIR} in order to maximize the retrieval rate. 

Following the notations in \cite{KarimAsymmetricPIR}, we denote the number of side information symbols that are used simultaneously in the initial round of downloads at the $n$th database by $s_n \in \{0, 1, \cdots, M-1\}$, e.g., if $s_n=1$, then the user requests to download a sum of 1 desired symbol and 1 undesired symbol as a side information in the form of $a+b$, $a+c$, ... etc., while $s_n=2$ implies that the user mixes every two undesired symbols to form one side information symbol, i.e., the user requests to download $a+b+c$, $a+c+d$, ... etc. For a given non-decreasing sequence $\{n_i\}_{i=0}^{M-1} \subset \{1, \cdots, N\}^M$, the databases are divided into groups, such that group 0 contains database 1 through database $n_0$, group 1 contains $n_1-n_0$ databases starting from database $n_0+1$, and so on.

Hence, let $s_n=i$ for all $n_{i-1}+1 \leq n \leq n_i$ with $n_{-1}=0$ by convention. Denote $\cs=\{i: s_n=i \:\text{for some}\: n \in \{1, \cdots, N\} \}$. We follow the round and stage definitions in \cite{MPIRjournal}. The $k$th round is the download queries that admit a sum of $k$ different messages ($k$-sum in \cite{JafarPIR}). A stage of the $k$th round is a query block of the $k$th round that exhausts all $\binom{M}{k}$ combinations of the $k$-sum. Denote $y_\ell[k]$ to be the number of stages in round $k$ downloaded from the $n$th database, such that $n_{\ell-1}+1 \leq n \leq n_\ell$. Our scheme is repeated for $\nu$ repetitions. Each repetition has the same query structure and operates over a block of message symbols of length $L^*$. Denote the total requested symbols from the $n$th database in one repetition of the scheme by $D_n(\mathbf{n})$. The details of the achievable scheme are as follows:

\begin{enumerate}
	\item \emph{Codebook construction:} According to the optimal probability distribution $p^*(x_n)$ (that maximizes the mutual information $I(X_n;Y_n)$), the $n$th database constructs a $\left(2^{\nu D_n(\mathbf{n})}, t_n(\mathbf{n})\right)$ codebook $\mathcal{C}_n$ at random, i.e., $p(x_{n,1}, \cdots, x_{n,t_n(\mathbf{n})})=\prod_{\eta_n=1}^{t_n(\mathbf{n})} p^*(x_{n,\eta_n})$. Specifically, the codebook $\mathcal{C}_n$ can be written as:
	\begin{align}
	\mathcal{C}_n=\begin{bmatrix}
	x_1(1) & x_2(1) & \cdots & x_{t_n(\mathbf{n})}(1) \\
	x_1(2) & x_2(2) & \cdots & x_{t_n(\mathbf{n})}(2) \\
	\vdots & \vdots & \vdots & \vdots \\
	x_1(2^{\nu D_n(\mathbf{n})}) & x_2(2^{\nu D_n(\mathbf{n})}) & \cdots & x_{t_n(\mathbf{n})}(2^{\nu D_n(\mathbf{n})})
	\end{bmatrix}_{2^{\nu D_n(\mathbf{n})} \times t_n(\mathbf{n})}
	\end{align}
	where 
	\begin{align}\label{answer_length}
	t_n(\mathbf{n})=\left\lceil \frac{\nu D_n(\mathbf{n})}{C_n}\right\rceil
	\end{align}
	This ensures that the rate of $\mathcal{C}_n$, $\frac{\nu D_n(\mathbf{n})}{t_n(\mathbf{n})} < C_n$ to ensure reliable transmission over the noisy channel. The $n$th database reveals the codebook $\mathcal{C}_n$ to the user.
	
	\item \emph{Initialization at the user side:} The user permutes each message independently and uniformly using a random interleaver, i.e., 
	\begin{align}
	\omega_m(i)=W_m(\pi_m(i)), \quad i \in \{1, \cdots, L\}
	\end{align}
	where $\omega_m(i)$ is the $i$th symbol of the permuted $W_m$, $\pi_m(\cdot)$ is a random interleaver for the $m$th message that is chosen independently, uniformly, and privately at the user's side. 
	\item \emph{Initial download:} From the $n$th database where $1 \leq n \leq n_0$, the user requests to download $\prod_{s \in \cs} \binom{M-2}{s-1}$ symbols from the desired message. The user sets the round index $k=1$. I.e., the user requests the desired symbols from $y_0[1]=\prod_{s \in \cs} \binom{M-2}{s-1}$ different stages.
	
	\item \emph{Message symmetry:} To satisfy the privacy constraint, for each stage initiated in the previous step, the user completes the stage by requesting the remaining $\binom{M-1}{k-1}$ $k$-sum combinations that do not include the desired symbols, in particular, if $k=1$, the user requests $\prod_{s \in \cs} \binom{M-2}{s-1}$ individual symbols from each undesired message.
	
	\item \emph{Database symmetry:} We divide the databases into groups. Group $\ell \in \cs$ corresponds to databases $n_{\ell-1}+1$ to $n_{\ell}$. Database symmetry is applied within each group only. Consequently, the user repeats step~2 over each group of databases, in particular, if $k=1$, the user downloads $\prod_{s \in \cs} \binom{M-2}{s-1}$ individual symbols from each message from the first $n_0$ databases (group 1).
	
	\item \emph{Exploitation of side information:} The undesired symbols  downloaded within the $k$th round (the $k$-sums that do not include the desired message) are used as side information in the $(k+1)$th round. This exploitation of side information is performed by requesting to download $(k+1)$-sum consisting of 1 desired symbol and a $k$-sum of undesired symbols only that were generated in the $k$th round. Note that for the $n$th database, if $s_n>k$, then this database does not exploit the side information generated in the $k$th round. Consequently, the $n$th database belonging to the $\ell$th group exploits the side information generated in the $k$th round from all databases except itself if $s_n \leq k$. Moreover, for $s_n=k$, extra side information can be used in the $n$th database. This is due to the fact that the user can form $n_0\prod_{s \in \cs\setminus \{s_n\}} \binom{M-2}{s-1}$ extra stages of side information by constructing $k$-sums of the undesired symbols in round 1 from the databases in group 0. 
	
	\item \emph{Repeat} steps 4, 5, 6 after setting $k=k+1$ until $k=M$. 
	
	\item \emph{Repetition of the scheme:} Repeat steps $3,\cdots,7$ for a total of $\nu$ repetitions. 
	
	\item \emph{Shuffling the order of the queries:} By shuffling the order of the queries uniformly, all possible queries can be made equally likely regardless of the message index. This guarantees the privacy.
	
	\item \emph{Encoding the responses to the user's queries:} The $n$th database responds to the user queries truthfully. The $n$th database concatenates all the responses to the user's queries in a vector $U_n^{[i]}$ of length $\nu D_n(\mathbf{n})$. The $n$th database uses $U_n^{[i]}$ as an index for choosing a codeword from $\mathcal{C}_n$, i.e., the index of the codeword and $U_n^{[i]}$ should be in bijection (e.g., by transforming $U_n^{[i]}$ into a decimal value). Consequently, the $n$th database responds with,
	\begin{align}
	A_n^{[i]}=[x_1(U_n^{[i]}) \quad x_1(U_n^{[i]}) \quad \cdots \quad x_{t_n(\mathbf{n})}(U_n^{[i]})]^T
	\end{align}

\end{enumerate}

\subsection{Privacy, Reliability, and Achievable Rate}
\paragraph{Privacy:} The privacy of the scheme follows from the privacy of the inherent PIR scheme under asymmetric traffic constraints. Specifically, for every stage of the $k$th round initiated, all $\binom{M}{k}$ combinations of the $k$-sum are included at each round. Thus, the structure of the queries is the same for any desired message at any repetition of the achievable scheme. Due to the random and independent permutation of each message and the random shuffling of the order of the queries, all queries are equally likely independent of the desired message index, and thus the privacy constraint in \eqref{privacy_constraint} is guaranteed.

\paragraph{Reliability:} The user employs \emph{joint typicality decoder} for every noisy answer string $\tilde{A}_n^{[i]}$ to decode the codeword index. From the channel coding theorem \cite[Theorem~7.7.1]{cover}, for every rate $\frac{\nu D_n(\mathbf{n})}{t_n(\mathbf{n})} < C_n$, there exists a sequence of $(2^{\nu D_n(\mathbf{n})}, t_n(\mathbf{n}))$ with maximum probability of error $P_e(\mathcal{C}_n) \rightarrow 0$ as $t_n(\mathbf{n}) \rightarrow \infty$. By letting $\nu \rightarrow \infty$, we have $t_n(\mathbf{n}) \rightarrow \infty$, $\frac{\nu D_n(\mathbf{n})}{t_n(\mathbf{n})} < C_n$ and hence we ensure the existence of a good code such that $P_e(\mathcal{C}_n) \rightarrow 0$. By union bound, the probability of error in decoding the indices of the codewords from every database is upper bounded by $P_e \leq \sum_{n=1}^{N} P_e(\mathcal{C}_n) \rightarrow 0$.

Since the index of the codeword is bijective to $U_n^{[i]}$, the probability of error in decoding $U_n^{[i]}$ for $n=1, \cdots, N$ is vanishingly small. Now, by construction of the queries as in \cite{KarimAsymmetricPIR}, all side information symbols used in the $(k+1)$th round are decodable in the $k$th round or from round 1, the user cancels out these side information and is left with symbols from the desired message. Consequently, there is no error in the decoding given that $U_n^{[i]}$ is correct for every $n$.

\paragraph{Achievable Rate:}  The structure of one repetition of our scheme is exactly as \cite{KarimAsymmetricPIR}. The recursive structure is described using the following system of difference equations that relate the number of stages in the databases belonging to a specific group as shown in \cite[Theorem~2]{KarimAsymmetricPIR}:
\begin{align}\label{difference_eqn}
y_0[k]&=(n_0\!-\!1)y_0[k\!-\!1]+\sum_{j \in \cs \setminus \{0\}} (n_j\!-\!n_{j-1}) y_j[k\!-\!1] \notag\\
y_1[k]&=(n_1\!-\!n_0\!-\!1)y_1[k\!-\!1]+\sum_{j \in \cs \setminus \{1\}} (n_j\!-\!n_{j-1}) y_j[k\!-\!1] \notag\\
y_\ell[k]&=n_0 \xi_\ell \delta[k\!-\!\ell\!-\!1]+(n_\ell\!-\!n_{\ell-1}\!-\!1) y_\ell[k-1]+\sum_{j \in \cs \setminus \{\ell\}} (n_j\!-\!n_{j-1})y_j[k\!-\!1], \quad  \ell \geq 2
\end{align}
where $y_\ell[k]$ is the number of stages in the $k$th round in a database belonging to the $\ell$th group, i.e., for the $n$th database, such that $n_{\ell-1}+1 \leq n \leq n_\ell$.

To calculate $D_n(\mathbf{n})$ where $n_{\ell-1} \leq n \leq n_{\ell}$, we note that for any stage in the $k$th round, the user downloads $\binom{M-1}{k-1}$ desired symbols from a total of $\binom{M}{k}$ downloads. Therefore,
\begin{align}
D_n(\mathbf{n})=\sum_{k=1}^{M} \binom{M}{k} y_\ell[k], \quad n_{\ell-1} \leq n \leq n_{\ell} 
\end{align}

Thus, the total download $\sum_{n=1}^N t_n(\mathbf{n})$ from all databases from all repetitions is calculated by observing \eqref{answer_length} and ignoring the ceiling operator as $\nu \rightarrow \infty$,
\begin{align}
\sum_{n=1}^N t_n(\mathbf{n})&= \sum_{n=1}^N\frac{\nu D_n(\mathbf{n})}{C_n}\\
&=\nu\left[\sum_{n=1}^{n_0}\frac{\sum_{k=1}^{M} \binom{M}{k} y_0[k] }{C_n}+\sum_{n=n_0+1}^{n_1}\frac{\sum_{k=1}^{M} \binom{M}{k} y_1[k]}{C_n}+\cdots\right]\\
&= \nu\sum_{\ell \in \cs} \sum_{n=n_{\ell-1}+1}^{n_\ell}\frac{\sum_{k=1}^{M} \binom{M}{k} y_\ell[k] }{C_n}
\end{align}
Furthermore, the total desired symbols from all databases from all repetitions is given by,
\begin{align}
L(\mathbf{n})=\nu\sum_{\ell \in \cs} \sum_{k=1}^{M}\binom{M-1}{k-1} y_\ell[k](n_\ell-n_{\ell-1})
\end{align}
Consequently, the following rate is achievable corresponding to the sequence $\mathbf{n}$,
\begin{align}
R(\mathbf{n},\bc)=\frac{\sum_{\ell \in \cs} \sum_{k=1}^{M}\binom{M-1}{k-1} y_\ell[k](n_\ell-n_{\ell-1})}{\sum_{\ell \in \cs} \sum_{n=n_{\ell-1}+1}^{n_\ell}\frac{\sum_{k=1}^{M} \binom{M}{k} y_\ell[k] }{C_n}}
\end{align}

Since this scheme is achievable for every monotone non-decreasing sequence $\mathbf{n}=\{n_i\}_{i=0}^{M-1}$, the following rate is achievable,
\begin{align}
R(\bc)= \max_{n_0 \leq \cdots \leq n_{M-1} \in \{1, \cdots, N\}} \frac{\sum_{\ell \in \cs} \sum_{k=1}^{M}\binom{M-1}{k-1} y_\ell[k](n_\ell-n_{\ell-1})}{\sum_{\ell \in \cs} \sum_{n=n_{\ell-1}+1}^{n_\ell}\frac{\sum_{k=1}^{M} \binom{M}{k} y_\ell[k] }{C_n}}
\end{align} 
   
\section{PIR from Multiple Access Channel}\label{MAC_PIR_sec}
\begin{figure}[t]
	\centering
	\includegraphics[width=1\textwidth]{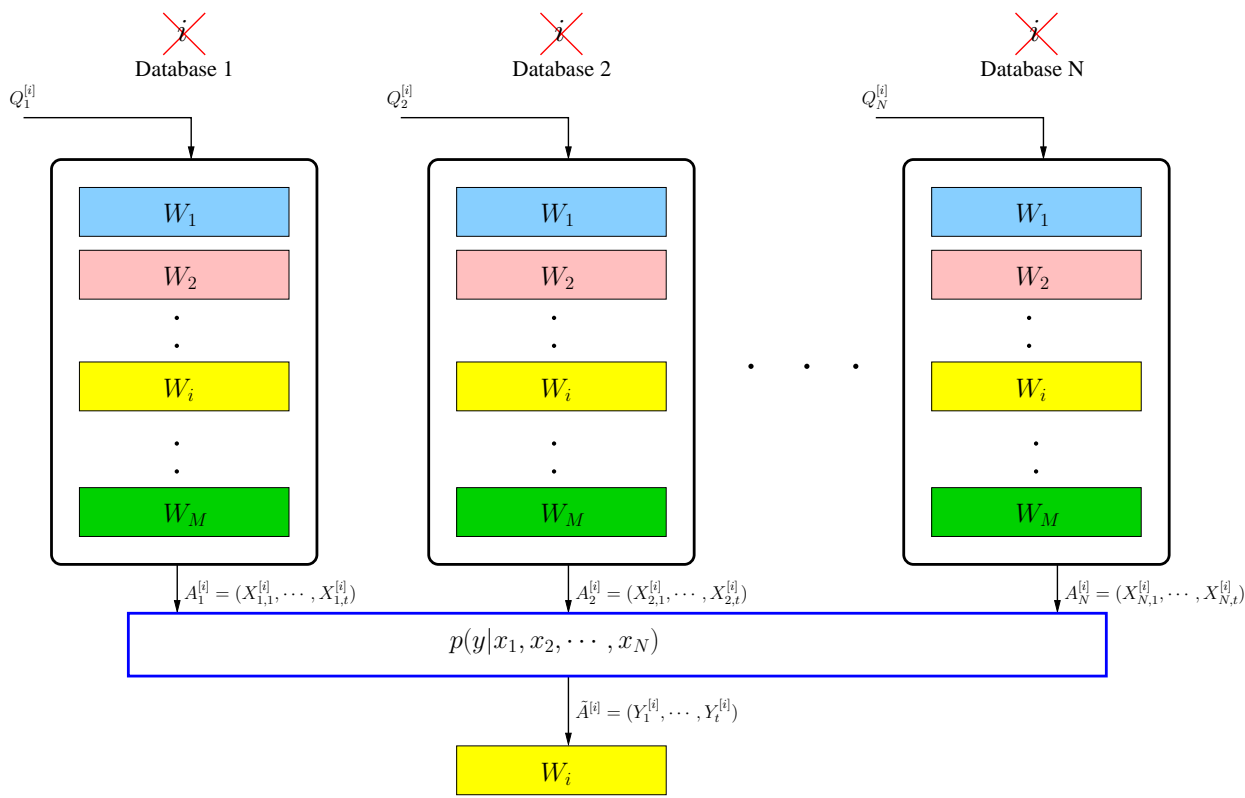}
	\caption{The MAC-PIR problem.}
	\label{MAC_PIR}
	\vspace*{-0.4cm}
\end{figure}

In this section, we consider the MAC-PIR problem. This problem is an extension of the NPIR model presented in Section~\ref{classicalPIR} which consists of $N$ non-colluding and replicated databases storing $M$ messages. In MAC-PIR (see Fig.~\ref{MAC_PIR}), the user sends a query $Q_n^{[i]}$ for the $n$th database to retrieve $W_i$ privately and correctly. The $n$th database responds with an answer string $A_n^{[i]}=(X_{n,1}^{[i]}, \cdots, X_{n,t}^{[i]})$. The user receives a noisy observation $\tilde{A}_n^{[i]}=(Y_1^{[i]}, \cdots, Y_t^{[i]})$, where the responses of the databases $(A_1^{[i]}, A_2^{[i]}, \cdots, A_N^{[i]})$ pass through a discrete memoryless channel with a transition probability distribution $p(y|x_1, \cdots, x_N)$, i.e., 
\begin{align}\label{MACmemoryless}
P\left(\tilde{A}^{[i]}|A_1^{[i]}, A_2^{[i]}, \cdots, A_N^{[i]} \right)=\prod_{\eta=1}^{t} p\left(y_{\eta}^{[i]}|x_{1,\eta}^{[i]}, x_{2,\eta}^{[i]}, \cdots, x_{N,\eta}^{[i]}\right)
\end{align}

In this sense, the retrieval is performed via a \emph{cooperative multiple access channel}, as the databases cooperate to convey the message $W_i$ to a common receiver (the user). The full cooperation is realized via the user queries. Furthermore, in MAC-PIR, the database responses are mixed together to have the noisy observation $\tilde{A}^{[i]}$ in contrast to the noisy PIR problem with orthogonal links presented in Section~\ref{classicalPIR}.

In MAC-PIR, the user should be able to reconstruct $W_i$ with vanishingly small probability of error by observing the noisy and mixed output $\tilde{A}^{[i]}$, i.e., the reliability constraint is written as:
\begin{align}\label{reliability2}
H(W_i|Q_{1:N}^{[i]}, \tilde{A}^{[i]}) \leq o(L)
\end{align} 
and the privacy constraint is written as:
\begin{align}\label{privacy2}
(Q_n^{[i]},A_n^{[i]},W_{1:M}) \sim (Q_n^{[j]},A_n^{[j]},W_{1:M}), \quad \forall i,j \in \{1, \cdots, M\} 
\end{align} 
We observe that the main difference between \eqref{privacy2} and \eqref{privacy_constraint} is that we cannot claim that $\tilde{A}^{[i]} \sim \tilde{A}^{[j]}$ in the MAC-PIR problem. This is due to the fact that the user cannot statistically differentiate between the responses corresponding to each message and hence the user cannot decode the desired message. This is in contrast to the NPIR problem with orthogonal links, where $\tilde{A}^{[i]} \sim \tilde{A}^{[j]}$ due to the Markov chain  $(W_{1:M},Q_{n}^{[i]}) \rightarrow A_n^{[i]} \rightarrow \tilde{A}_n^{[i]}$.

The retrieval rate for the MAC-PIR is given by:
\begin{align}
R=\frac{L}{t}
\end{align}
and the MAC-PIR capacity is $C_{\text{PIR}}=\sup \: R$ over all retrieval schemes. We note that, without loss of generality, we can assume that all responses from the databases have the same length $t$ in contrast to the NPIR problem with orthogonal links. The reason is that the retrieval rate depends only on the output of the channel and not on the individual responses of the databases. Hence, even if the database responses are different in lengths, we can choose $t=\max_{n \in [N]} \: t_n$ by appending the remaining responses by dummy symbols.

In the sequel, we discuss the issue of separability of channel coding and the information retrieval in MAC-PIR via some examples. Interestingly, we show that the optimal PIR scheme for the additive MAC and logic conjunction/disjunction MAC, the channel coding and the retrieval scheme are dependent on the channel transition probability, and hence channel coding and retrieval procedure are inseparable. 

\subsection{Additive MAC}
In the first special case, we consider the additive MAC. In the additive MAC, at each time instant $\eta$, the responses of the databases are added together (in modulo-2) in addition to a random variable $Z_\eta \sim \text{Bernoulli}(p)$, which is independent of $(W_{1:M},Q_{1:N}^{[i]})$ and corresponds to a random additive noise, i.e., 
\begin{align}
Y_\eta=\sum_{n=1}^N X_{n,\eta}+Z_\eta
\end{align}

The following theorem characterizes the capacity of the MAC-PIR problem if the channel is restricted to additive MACs.

\begin{theorem}
	The additive MAC-PIR capacity is,
	\begin{align}
	C_\text{PIR}=1-H(p)
	\end{align}
	where $p \in [0,0.5)$ is the flipping probability of the additive noise.
\end{theorem}

We have the following remarks.
\begin{remark}
	For noiseless additive MAC, i.e., $p=0$ and $Y_\eta=\sum_{n=1}^N X_{n,\eta}$, the MAC-PIR capacity is $C_\text{PIR}=1$. This implies that there is no penalty due to the privacy constraint, i.e., the user can have privacy for free. Interestingly, this is the first instance where the PIR capacity is independent of the number of databases $N$ and the number of messages $M$.
\end{remark}

\begin{remark}
	For noiseless additive MAC, i.e., $p=0$, separation between channel coding and retrieval process is not optimal unlike the NPIR problem with orthogonal links. In fact, the retrieval scheme is dependent on the structure of the channel. To see this, the user generates a random binary vector $\mathbf{h}=[h_1 \: h_2 \: \cdots \: h_M] \in \{0,1\}^M$.  The user sends $\mathbf{h}$ to database~1, flips the $i$th position of $\mathbf{h}$ and sends it to database~2, and does not send anything to the remaining databases. Thus, the responses of the databases are,
	\begin{align}
	A_1^{[i]}&=\sum_{m=1}^M h_m W_m \\
	A_2^{[i]}&=\sum_{m=1}^M h_m W_m+W_i 
	\end{align}
	This is exactly the retrieval scheme in \cite{ChorPIR}. Since the channel is additive and noiseless, $\tilde{A}^{[i]}=A_1^{[i]}+A_2^{[i]}=W_i$. Hence, the user downloads 1 bit from the channel in order to get 1 bit from the desired file and $R=1$. Here, we note that, the channel performs the processing at the user for free. This implies that by careful design of queries, the user can exploit the channel in its favor to maximize the retrieval rate.
\end{remark}

\begin{Proof}
	We prove the converse and achievability.
	
	\paragraph{The converse proof:} To show the converse, we assume that $W_1$ is the desired message without loss of generality. Then, we have the following implications, 
	\begin{align}
	L&\:\:\:=H(W_1) \\
	 &\stackrel{\eqref{msg_indep},\eqref{independency}}{=}H(W_1|W_{2:M},Q_{1:N}^{[1]})\label{thm31} \\
	 &\:\:\stackrel{\eqref{reliability2}}{\leq}H(W_1|W_{2:M},Q_{1:N}^{[1]})-H(W_1|W_{2:M},Q_{1:N}^{[1]},\tilde{A}^{[1]})+o(L) \label{thm32}\\
	 &\:\:\: =I(W_1;\tilde{A}^{[1]}|Q_{1:N}^{[1]},W_{2:M})+o(L) \\
	 &\:\:\: =H(\tilde{A}^{[1]}|Q_{1:N}^{[1]},W_{2:M})-H(\tilde{A}^{[1]}|Q_{1:N}^{[1]},W_{1:M})+o(L)\\
	 &\:\:\stackrel{\eqref{answer_constraint}}{\leq} H(\tilde{A}^{[1]})-H(\tilde{A}^{[1]}|Q_{1:N}^{[1]},W_{1:M},A_{1:N}^{[1]})+o(L) \label{thm33}\\
	 &\:\:\: =t-H(\tilde{A}^{[1]}|A_{1:N}^{[1]})+o(L) \label{thm34}\\
	 &\:\:\: =t-\sum_{\eta=1}^{t}H(Y_\eta^{[1]}|X_{1,\eta}^{[1]},X_{2,\eta}^{[1]},\cdots,X_{N,\eta}^{[1]})+o(L) \label{thm35}\\ 
	 &\:\:\: =t-\sum_{\eta=1}^{t}H\left(\sum_{n=1}^N X_{n,\eta}^{[1]}+Z_\eta|X_{1,\eta}^{[1]},X_{2,\eta}^{[1]},\cdots,X_{N,\eta}^{[1]}\right)+o(L)\\
	 &\:\:\: =t-\sum_{\eta=1}^{t} H(Z_\eta|X_{1,\eta}^{[1]},X_{2,\eta}^{[1]},\cdots,X_{N,\eta}^{[1]})+o(L) \\
	 &\:\:\: =t(1-H(p))+o(L) \label{thm36}
	\end{align}
	where \eqref{thm31} follows from the independence of the messages and the queries, \eqref{thm32} follows from the reliability constraint, \eqref{thm33} follows from the fact that the answer string $A_n^{[1]}$ is a deterministic function of the messages and the queries, \eqref{thm34} follows from the fact that $(W_{1:M},Q_{1:N}^{[1]}) \rightarrow A_{1:N}^{[1]} \rightarrow \tilde{A}^{[1]}$ is a Markov chain, \eqref{thm35} follows from the fact that the channel is memoryless, and \eqref{thm36} follows from the independence of $Z_\eta$ and $(X_{1,\eta}^{[1]},X_{2,\eta}^{[1]},\cdots,X_{N,\eta}^{[1]})$ as a consequence of the independence of $(Z_\eta, W_{1:M}, Q_{1:N}^{[1]})$.
	
	Hence, by reordering terms and taking $L \rightarrow \infty$, we have $R= \frac{L}{t} \leq 1-H(p)$. Note that we can interpret the upper bound as the cooperative MAC bound, i.e., $R \leq I(Y; X_1,X_2, \cdots, X_N)=1-H(p)$.
	
	\paragraph{The achievability proof:} To show the general achievability, the user submits queries to database~1 and database~2 only and ignores the remaining databases. We note that the additive MAC in this case boils down to $Y_\eta=X_{1,\eta}+X_{2,\eta}+Z_\eta$, which means that the channel $p(y|x_1,x_2)$ is BSC($p$). Consequently, we use again Shannon's coding theorem for BSC in Lemma~\ref{lemma_BSC}.
	
	To that end, let the $m$th message be a vector $W_m=[W_{m}(1) \quad W_{m}(2) \quad \cdots \quad W_{m}(L)]$ of length $L$. The user repeats the following scheme $L$ times. For the $j$th repetition of the scheme, the user generates a random binary vector $\mathbf{h}(j)=[h_1(j) \quad h_2(j) \quad \cdots\quad h_M(j)] \in \{0,1\}^M$. The user sends the following queries to the databases:
	\begin{align}
	Q_1^{[i]}(j)&=\mathbf{h}(j) \\
	Q_2^{[i]}(j)&=\mathbf{h}(j)+\mathbf{e}_i
	\end{align}
	where $\mathbf{e}_i$ is the unit vector containing 1 only at the $i$th position. The queries are private since $Q_n^{[i]}$ is a vector picked uniformly from $\{0,1\}^M$ for any message $i$.
	
	For the $j$th repetition of the scheme, the database uses the received query vector as a combining vector for the $j$th element of all messages. The $n$th database concatenates all responses in a vector $U_n^{[i]}$ of length $L$, hence
	\begin{align}
	U_1^{[i]}&=\left[\sum_{m=1}^M h_m(1) W_m(1) \quad \sum_{m=1}^M h_m(2) W_m(2) \quad \cdots \quad \sum_{m=1}^M h_m(L) W_m(L) \right] \\
	U_2^{[i]}&=\left[\sum_{m=1}^M h_m(1) W_m(1)+W_i(1) \quad \sum_{m=1}^M h_m(2) W_m(2)+W_i(2) \right.\notag\\
	         &\quad\quad \cdots \quad \left.\sum_{m=1}^M h_m(L) W_m(L)+W_i(L) \right]
	\end{align}
	
	From Lemma~\ref{lemma_BSC}, for $p \in (0,0.5)$, all but $\rho$ linear $[t,L]$ block codes  $\mathcal{C}$, where $\frac{L}{t}=R<1-H(p)$ that have  $P_e(\mathcal{C}) < \frac{2}{\rho} \cdot 2^{-t\Delta(p,R)}$. Then, the databases agree on the same $[t,L]$ code from the family of good codes, where $t=\frac{L}{\lfloor1-H(p)\rfloor}$. The $n$th database encodes $U_n^{[i]}$ independently by the same $[t,L]$ linear block code to output $A_n^{[i]}$. 
	
	After passing through the noisy channel, the noisy observation is given by:
	\begin{align}
	\tilde{A}^{[i]}&=A_1^{[i]}+A_2^{[i]}+Z_{1:t}\\
	               &=\hat{A}^{[i]}+Z_{1:t} \label{MAC_BSC}
	\end{align}
	Since the two databases employ the same linear block code, the sum of the two codewords $\hat{A}^{[i]}=A_1^{[i]}+A_2^{[i]}$ is also a valid codeword corresponding to the sum $U_1^{[i]}+U_2^{[i]}$.
	
	Consequently, as $L \rightarrow \infty$, $t \rightarrow \infty$, the probability of error in decoding the sum $U_1^{[i]}+U_2^{[i]}$ is $P_e(L) \rightarrow 0$. By observing that $U_1^{[i]}+U_2^{[i]}=W_i$, the reliability proof follows.	
\end{Proof}

\begin{remark}
	In the achievability proof, the PIR scheme relies on the additivity of the channel. In particular, the scheme uses a linear block code to exploit the fact that the sum of two codewords from a linear block code is also a valid codeword. Consequently, the retrieval process depends on the channel transition probability explicitly as opposed to the NPIR problem with orthogonal links.
\end{remark}

\subsection{Logic Conjunction/Disjunction MACs}
In this section, we show that we can achieve privacy for free for MACs other than the additive MACs. We illustrate this result by considering the MAC-PIR problem through channels that output the logical conjunctions (logic AND)/disjunctions (logic OR) of the inputs. Let $\wedge$ denote the logical conjunction operator, $\vee$ denote the logical disjunction operator, and $\neg$ denote the logical negation operator. The input-output relation of the discrete memoryless logical conjunction channel is given as:
\begin{align}
Y_\eta=\bigwedge_{n=1}^N X_{n,\eta}
\end{align}

For the logical conjunction channel, we have the following capacity result.
\begin{theorem}
	In the logical conjunction MAC-PIR problem, if $N \geq 2^{M-1}$, then the MAC-PIR capacity is $C_{\text{PIR}}=1$, where $M$ is the number of messages.
\end{theorem}

We have the following observations:
\begin{remark}
	Similar to the additive MAC, there is no loss due to the privacy constraint for the conjunction MAC. In this case, the capacity depends on the number of messages $M$, and the number of databases $N$ unlike the additive MAC. Interestingly, the result shows the first instance of a threshold for the number of databases at which the full unconstrained capacity can be achieved $N=2^{M-1}$, which is dependent on the number of messages $M$.   
\end{remark}

\begin{remark}
	We note that the minimum number of databases $N$ that results in $C_{\text{PIR}}=1$ is still an open problem. In fact, the capacity for $N < 2^{M-1}$ is also an interesting open problem. 
\end{remark}

\begin{Proof}
	It suffices to show only the achievability for this problem as the retrieval rate is trivially upper bounded by 1. To that end, the user submits queries to $2^{M-1}$ databases and submits nothing to the remaining databases. The user generates the random variables $(Z_1, \cdots, Z_M)$ independently, privately, and uniformly from $\{0,1\}$. The random variable $Z_m \sim \text{Bernoulli}(\frac{1}{2})$ is a Bernoulli random variable that represents the negation state of the $m$th message literal in the first query $Q_1^{[i]}$, i.e., if $Z_m=1$, this means that the user requests $W_m$ in $Q_1^{[i]}$, while $Z_m=0$ means that the user requests $\neg W_m$ in $Q_1^{[i]}$. Let $\tilde{W}_m$ be the requested literal from the $m$th message in $Q_1^{[i]}$, hence,
	\begin{align}
	\tilde{W}_m=\left\{
	\begin{array}{ll}
	W_m, \quad &Z_m=1 \\
	\neg W_m, \quad &Z_m=0
	\end{array}\right.
	\end{align}
	
	Now, without loss of generality, assume that $W_1$ is the desired message. From database~1, the user requests to download the disjunction $X_1=\bigvee_{m=1}^M \tilde{W}_m$. From every other database, the user requests the same literal $\tilde{W}_1$ with a new disjunction of the remaining messages with different negation pattern than what is requested from database~1. I.e., from database~2, the user requests the disjunction $X_2=\tilde{W}_1 \vee \neg \tilde{W}_2 \vee \bigvee_{m \in [M]\setminus\{1,2\}} \tilde{W}_m$. From database~3, the user requests the disjunction $X_3=\tilde{W}_1 \vee \neg \tilde{W}_3 \vee \bigvee_{m \in [M]\setminus\{1,3\}} \tilde{W}_m$, $\cdots$ etc. Denote the disjunction of messages $W_{2:M}$ requested from the $n$th database by $F_n$, where $n \in \{1, \cdots, 2^{M-1}\}$, then the received observation at the user is 
	\begin{align}
	Y&=\left(\bigvee_{m=1}^M \tilde{W}_m\right) \wedge \left(\tilde{W}_1 \vee \neg \tilde{W}_2 \vee\!\!\!\! \bigvee_{m \in [M]\setminus\{1,2\}} \!\!\!\!\tilde{W}_m\right) \wedge \left(\tilde{W}_1 \vee \neg \tilde{W}_3 \vee \!\!\!\! \bigvee_{m \in [M]\setminus\{1,3\}} \!\!\!\! \tilde{W}_m\right) \wedge \cdots \\
	 &=\tilde{W}_1 \vee \bigwedge_{i=1}^{2^{M-1}} F_i \label{conjProof1}\\
	 &=\tilde{W}_1 \label{conjProof2}
	\end{align} 
	where \eqref{conjProof1} follows from successively applying the Boolean relation $(\tilde{W}_1 \vee G_1) \wedge (\tilde{W}_1 \vee G_2)=\tilde{W}_1 \vee (G_1 \wedge G_2)$ for any logical expressions $G_1$, $G_2$. \eqref{conjProof2} follows from the fact that there exist $2^{M-1}$ different negation states for the literals from $W_{2:M}$, each negation state is requested from one database in the form of logical expression $F_i$, hence the conjunction of all these logical expressions $\bigwedge_{i=1}^{2^{M-1}} F_i=0$ as all possible product of sums of $W_{2:M}$ exist in the conjunction. This satisfies the reliability constraint. Another way to see this result is that the queries are designed such that they cover \emph{exactly half} the $M$-dimensional Karnaugh map, which can be reduced to either $W_1$ or $\neg W_1$.
	
	Furthermore, since the negation state for every message is chosen uniformly, independently, and uniformly for each message, the probability of receiving specific query from the user is $\frac{1}{2^M}$ irrespective to the desired message, which guarantees the privacy.    
\end{Proof}

\paragraph{Illustrative example: $M=3$ messages, $N=4$ databases with conjunction channel:}
As an explicit example, let $M=3$, $N=2^{M-1}=4$, then the user requests the following:
\begin{align}
X_1&=\tilde{W}_1 \vee \tilde{W}_2 \vee \tilde{W}_3 \\
X_2&=\tilde{W}_1 \vee \neg \tilde{W}_2 \vee \tilde{W}_3\\
X_3&=\tilde{W}_1 \vee \tilde{W}_2 \vee \neg \tilde{W}_3\\
X_4&=\tilde{W}_1 \vee \neg \tilde{W}_2 \vee \neg \tilde{W}_3
\end{align}
Hence, the output of the channel is,
\begin{align}
Y&=X_1 \wedge X_2 \wedge X_3 \wedge X_4\\
&=(\tilde{W}_1 \vee \tilde{W}_2 \vee \tilde{W}_3)\wedge(\tilde{W}_1 \vee \neg \tilde{W}_2 \vee \tilde{W}_3)\wedge(\tilde{W}_1 \vee \tilde{W}_2 \vee \neg \tilde{W}_3)\wedge(\tilde{W}_1 \vee \neg \tilde{W}_2 \vee \neg \tilde{W}_3) \\
&=(\tilde{W}_1\vee(\tilde{W}_2 \vee \tilde{W}_3)\wedge(\neg \tilde{W}_2 \vee \tilde{W}_3))\wedge(\tilde{W}_1\vee(\tilde{W}_2 \vee \neg\tilde{W}_3)\wedge(\neg \tilde{W}_2 \vee \neg\tilde{W}_3)) \\
&=(\tilde{W}_1\vee W_3)\wedge(\tilde{W}_1 \vee \neg \tilde{W}_3) \\
&=\tilde{W}_1
\end{align}
Thus, the user can decode $W_1$ from $Y$ as the user knows the correct negation pattern for $\tilde{W}_1$ privately. The scheme is private as all queries are equally likely with probability $\frac{1}{8}$ irrespective to the desired message. Since the user downloads 1 bit to retrieve 1 bit from the desired message, the retrieval rate $R=1$.

\begin{remark}
	We note that the result is still valid if the channel is replaced by a disjunction channel, i.e.,
	\begin{align}
	Y_\eta=\bigvee_{n=1}^N X_{n,\eta}
	\end{align}
	In this case, the user submits the same queries for the databases with replacing every disjunction operator with a conjunction operator. The proof of reliability follows from the duality of the product-of-sum and the sum-of-product. 
\end{remark}

\begin{remark}
	The achievable scheme for the conjunction channel is a non-linear retrieval scheme that depends on the non-linear characteristics of the channel in contrast to the linear retrieval scheme used for the additive channel. This confirms the non-separability between the retrieval scheme and the channel coding needed for reliable communication through the channel. 
\end{remark}


\subsection{Selection Channel}
In this example, we illustrate the fact that the \emph{privacy for free} phenomenon may not be always feasible for any arbitrary channel in the MAC-PIR problem. To illustrate this, we consider the selection channel. In this channel, the user selects to connect to one database only at random and sticks to it throughout the transmission, i.e.,
\begin{align}
Y_\eta=X_{n,\eta}, \quad n \sim \text{uniform}\:\{1, \cdots, N\}
\end{align} 

In this channel, the user is connected to the same database at every channel use. This implies that the user faces a single-database ($N=1$) PIR problem at every channel use. The optimal PIR strategy for $N=1$ is to download all the messages ($M$ messages) from the connected database. Thus, the PIR capacity is given by $C_{\text{PIR}}=\frac{1}{M}$.

It is worth noting that there is another slight variant of the selection channel, in which the user selects to connect to one database at random at every channel use, i.e.,
\begin{align}
Y_\eta=X_{n(\eta),\eta}, \quad n(\eta) \sim \text{uniform}\:\{1, \cdots, N\}
\end{align}
where $n(\eta)$ corresponds to the database index at channel use $\eta$. Then, $C_{PIR} \leq C= (1+\frac{1}{N}+\cdots+\frac{1}{N^{M-1}})^{-1}$ trivially as the capacity of the classical PIR $C$ \cite{JafarPIR}, in which all the databases are connected to the user, is an upper bound for this problem, as the user can choose to ignore all the responses except the ones in the classical PIR problem. For the achievability, the user can repeat the achievable scheme in \cite{JafarPIR} $\nu$ times, which results in using the selection channel $t=\nu \frac{L}{C}=\nu \frac{N(N^M-1)}{N-1} $. At channel use $\eta$, the user chooses a new query element from $Q_{n(\eta)}^{[i]}$ and submits it to database $n(\eta)$. As $\nu \rightarrow \infty$, by strong law of large numbers, each database will be visited $t_n$ times, where $t_n \rightarrow \frac{t}{N}$ in the limit for every $n$. Hence, all bits are decodable by the decodability of the scheme in \cite{JafarPIR} and $C_{PIR}=C= (1+\frac{1}{N}+\cdots+\frac{1}{N^{M-1}})^{-1} < 1$ as well.

\section{Conclusion}
In this paper, we introduced noisy PIR with orthogonal links (NPIR), and PIR from multiple access channels (MAC-PIR). We focused on the issue of the separability of the channel coding and the retrieval scheme. For the NPIR problem, we proved that the channel coding and the retrieval scheme are \emph{almost separable} in the sense that every database implements its own channel coding independently from other databases. The problem is coupled only through agreeing on a suitable traffic ratio vector to maximize the retrieval rate. On the other hand, these conclusions are not valid for the MAC-PIR problem. We showed two examples, namely: PIR from additive MAC and PIR from logical conjunction/disjunction MAC. In these examples, we showed that the channel coding and retrieval schemes are indeed \emph{inseparable} unlike in the NPIR problem. In both cases, we showed that by careful design of joint retrieval and coding schemes, we can attain the full capacity $C_{PIR}=1-H(p)$ and $C_{PIR}=1$, respectively, with no loss due to the privacy constraint.

\bibliographystyle{unsrt}
\bibliography{references}
\end{document}